\documentclass[pdflatex,sn-nature]{sn-jnl}
\usepackage{natbib}
\setcitestyle{square, comma, numbers, sort&compress, super}
\usepackage{amsmath,amssymb}
\usepackage{manyfoot}
\usepackage{graphicx}
\usepackage{caption}
\usepackage{url}
\usepackage{setspace}
\usepackage{multibib}
\usepackage{booktabs}
\newcites{supp}{Methods References}

\usepackage{graphicx}	

\usepackage[T1]{fontenc} 
\usepackage{academicons}
\usepackage{xcolor}
\usepackage{multirow}
\usepackage{float}
\usepackage{subcaption}
\usepackage[superscript,biblabel]{cite}
\usepackage{url}
\usepackage{hyperref}
\usepackage{soul}
\usepackage{rotating}

\usepackage{newtxtext,newtxmath}

\newcommand{\HI}{\textrm{H\,{\sc i}}}
\newcommand{\CI}{\textrm{C\,{\sc i}}}
\newcommand{\SiII}{\textrm{Si\,{\sc ii}}}
\newcommand{\CII}{\textrm{C\,{\sc ii}}}
\newcommand{\OI}{\textrm{O\,{\sc i}}}
\newcommand{\FeII}{\textrm{Fe\,{\sc ii}}}
\newcommand{\AlII}{\textrm{Al\,{\sc ii}}}
\newcommand{\CIV}{\textrm{C\,{\sc iv}}}
\newcommand{\CIII}{\textrm{C\,{\sc iii}}}

\newcommand{\cmsq}{\mbox{\textrm{cm}$^{-2}$}}
\newcommand{\kms}{\mbox{\textrm{km}\,s$^{-1}$}}

\newcommand{\qso}{J\,012555.11$-$012925.00}

\newcommand{\araa}{Annu. Rev. Astron. Astrophys.} 
\newcommand{\aj}{Astron. J.} 
\newcommand{\apj}{Astrophys. J.} 
\newcommand{\apjl}{Astrophys. J. Lett.} 
\newcommand{\apjs}{Astrophys. J. Suppl. Ser.} 
\newcommand{\ao}{Appl. Opt.} 
\newcommand{\aap}{Astron. Astrophys.} 
\newcommand{\aapr}{Astron. Astrophys. Rev.} 
\newcommand{\mnras}{Mon. Not. R. Astron. Soc.} 
\newcommand{\nat}{Nature} 
\newcommand{\pasj}{Publ. Astron. Soc. Jpn} 
\newcommand{\pasp}{Publ. Astron. Soc. Pac.} 

\begin{document}

\title{Quasar radiation transforms the gas in a merging companion galaxy}

\author*[1]{\fnm{Sergei} \sur{Balashev}}\email{s.balashev@gmail.com}
\equalcont{These authors contributed equally to this work.}
\author*[2,3]{\fnm{Pasquier} \sur{Noterdaeme}}\email{noterdaeme@iap.fr}
\equalcont{These authors contributed equally to this work.}
\author[4]{\fnm{Neeraj} \sur{Gupta}}
\author[3,5]{\fnm{Jens-Kristian} \sur{Krogager}}
\author[6]{\fnm{Francoise} \sur{Combes}}
\author[7]{\fnm{Sebasti{\'a}n} \sur{L{\'o}pez}}
\author[2]{\fnm{Patrick} \sur{Petitjean}}
\author[2]{\fnm{Alain} \sur{Omont}}
\author[4]{\fnm{Raghunathan} \sur{Srianand}}
\author[7]{\fnm{Rodrigo} \sur{Cuellar}}

\affil[1]{\orgdiv{Department of Theoretical Astrophysics}, \orgname{Ioffe Institute}, \orgaddress{\street{Politechnicheskaya ul., 26}, \city{Saint-Peterburg}, \postcode{194021}, \country{Russia}}}
\affil[2]{\orgname{Institut d'Astrophysique de Paris, CNRS-SU, UMR\,7095}, \orgaddress{\street{98bis~bd~Arago}, \postcode{75014} \city{Paris}, \country{France}}}
\affil[3]{\orgdiv{French-Chilean Laboratory for Astronomy}, \orgname{IRL\,3386, CNRS and Universidad de Chile}, \orgaddress{\street{Casilla 36-D}, \city{Santiago}, \country{Chile}}}
\affil[4]{\orgname{Inter-University Centre for Astronomy and Astrophysics}, \orgaddress{\street{Post Bag 4, Ganeshkhind, 411 007}, \city{Pune}, \country{India}}}
\affil[5]{\orgdiv{Centre de Recherche Astrophysique de Lyon, UMR\,5574}, \orgname{Université Lyon I, ENS de Lyon, CNRS}, \orgaddress{\street{F-69230 Saint-Genis-Laval}, \city{Lyon}, \country{France}}}
\affil[6]{\orgdiv{Collège de France, PSL University, Sorbonne University, CNRS, LERMA}, \orgname{Observatoire de Paris}, \orgaddress{\street{61 avenue de l’Observatoire}, \postcode{75014}, \city{Paris}, \country{France}}}
\affil[7]{\orgdiv{Departamento de Astronomía}, \orgname{Universidad de Chile}, \orgaddress{\street{Casilla 36-D}, \city{Santiago}, \country{Chile}}}
\maketitle

\textbf{
Quasars, powered by gas accretion onto supermassive black holes\cite{Antonucci1993,Netzer2015}, rank among the most energetic objects of the Universe\cite{Wu2015,Wolf2024}. 
While they are thought to be ignited by galaxy mergers\cite{Volonteri2003,Hopkins2008,Ellison2011,Trakhtenbrot2017,Decarli2017,Goulding2018,Fogasy2020} and affect the surrounding gas\cite{Hopkins2010,Dubois2016,Pontzen2017,Moiseev2023}, observational constraints on both processes remain scarce\cite{Fabian2012,Hirschmann2012,Tang2023}.
Here we unveil a major merging system at redshift $z \approx 2.7$, and demonstrate that radiation from the quasar in one galaxy directly alters the gas properties in the other galaxy.
Our findings reveal that the galaxies, with centroids separated by only a few kiloparsecs and approaching each other at speed $\approx550$\,\kms, are massive, form stars, and contain a substantial molecular mass. Yet, dusty molecular gas seen in absorption against the quasar nucleus is highly excited and confined within cloudlets with densities $\sim 10^5$ - $10^6$ cm$^{-3}$ and sizes $<$0.02 pc, several orders of magnitude more compact than those observed in intervening (non-quasar) environments. This is also approximately 10$^5$ times smaller than currently resolvable through molecular-line emission at high redshifts. We infer that, wherever exposed to the quasar radiation, molecular gas is disrupted, leaving behind surviving dense clouds too small to give birth to new stars.
Our results not only underscore the role of major galaxy mergers in triggering quasar activity, but also reveal localized negative feedback as a profound alteration of internal gas structure which likely hampers star formation. 
}
\bigskip

We combine unique millimeter and optical observations of the quasar \qso\ to investigate its galactic environment and directly assess the impact of its intense radiation on the physical state and small-scale structure of the molecular gas observed in absorption\cite{Noterdaeme2019}.
In Fig.~1, we show that even though the optical image (a) is dominated by the overwhelming quasar light, a careful subtraction of the quasar's point spread function reveals the presence of a companion galaxy and a structure extending around over more than 20~kpc in visible light (b).
Our high spatial resolution ALMA observations of the dense molecular component (d) reveal that the centroids of the two galaxies --quasar host and companion-- are separated by only $\sim$0.7 arcsec on the plane of the sky, that is, a physical projected distance of about 5~kpc at redshift $z_{\rm syst} = 2.6618$. The companion galaxy has a slightly higher apparent redshift than the quasar host-galaxy, with a line-of-sight velocity relative to the quasar, obtained from the CO(7-6) emission lines, of about 550~\kms (e, f and g). 

\begin{figure*}[!htbp]
\centering
\includegraphics[width=1.0\textwidth]{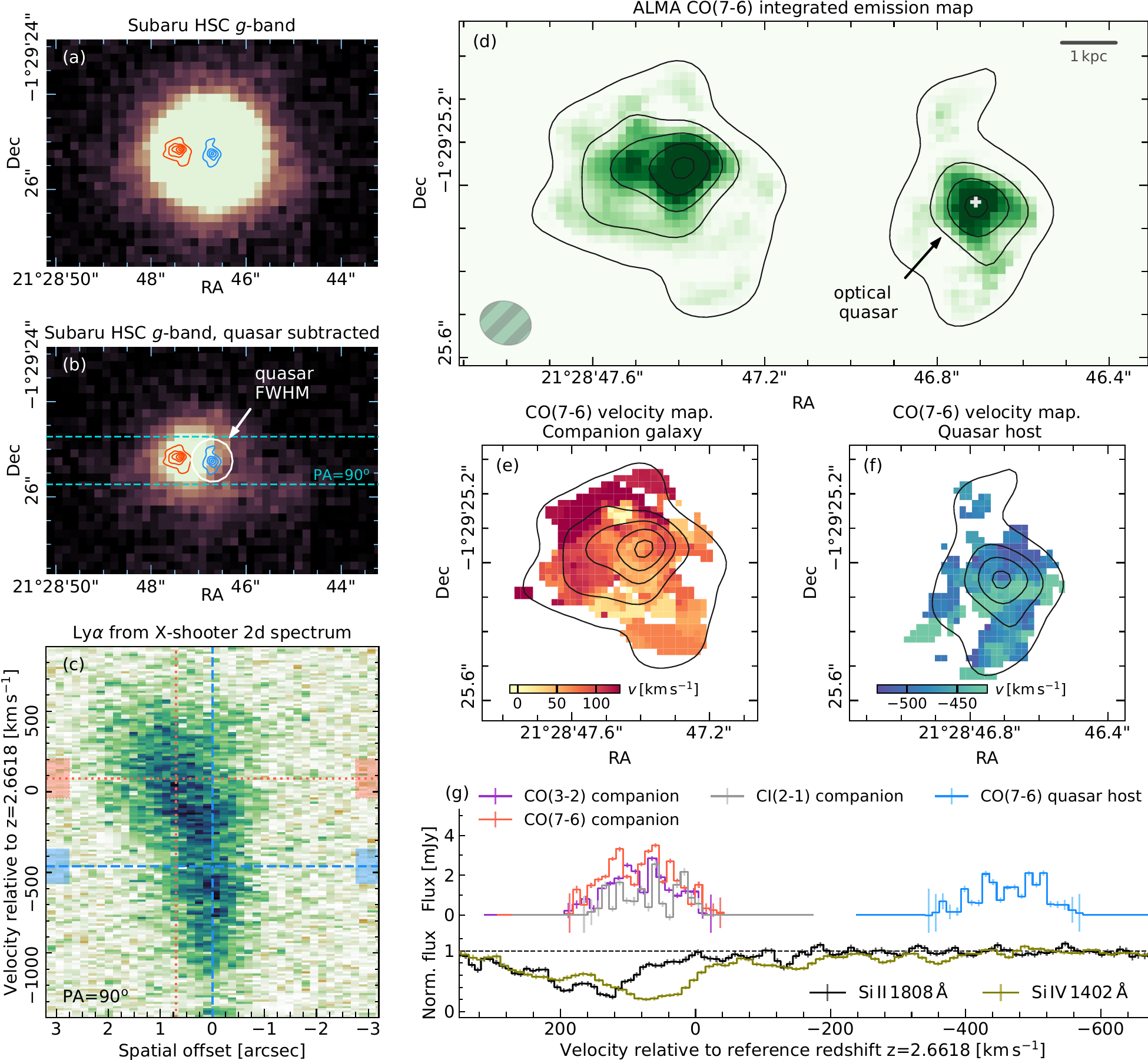}
\caption{
\textbf{Optical and millimeter observations of the quasar-host companion-galaxy system.}\\
\noindent
\textbf{a}, The Subaru HSC $g$-band image is dominated by the emission from the quasar but the companion galaxy is clearly seen $\sim 1$~arcsec to the East, after subtraction of the quasar point spread function (\textbf{b}). Contours from the ALMA CO(7-6) emission map (blue and red) are superimposed, as well as the layout of the X-shooter slits. \textbf{d}, The integrated CO(7-6) emission with ALMA, with contours from 5 with 20 mJy\,beam$^{-1}$\,km\,s$^{-1}$ step. The hatch ellipse depicts the beam size, the white cross represents the centroid of g-band emission from HSC. \textbf{e}, \textbf{f}, The velocity maps at the location of the companion and the quasar host, respectively. \textbf{c}, The 2D X-shooter spectrum with slit position angle=90$^\circ$, where the quasar light is naturally suppressed by strong absorption by \HI\ gas, providing direct access to the extended Ly-$\alpha$ emission from merging galaxies. The red dotted and blue dashed lines denote the centroids of the ALMA CO emission of the companion and quasar host galaxies, respectively, in both spatial and velocity space, with the stripes indicating their velocity spread.
\textbf{g}, The CO and CI integrated velocity profiles with the absorption profiles of metal lines in the 1D X-shooter spectrum of the quasar. The error bars represent the $1\sigma$ standard deviation of the flux measurements.
\label{fig:emission}
}
\end{figure*}

The use of different slit orientations during X-shooter observations allowed us to extract both the quasar (PA=0$^\circ$ and PA=90$^\circ$) and the companion galaxy (PA=90$^\circ$) {optical} spectra. 
A key observation is the presence of strong absorptions by ionised, neutral and molecular gas in the quasar spectrum, at the redshift of the companion galaxy (as shown in panel g in Fig.~1 and in Fig.~2). The companion galaxy must therefore be situated in the foreground, showing that the objects are also approaching each other, indicating a merging scenario. In addition, the absorption lines conveniently remove the otherwise outshining light from the quasar central engine, acting like a coronagraph which reveals emission from the host galaxy at various wavelengths. This is seen as a spatially resolved residual in the trough of saturated metal and molecular hydrogen absorption lines (see Methods). Remarkably, the damped Ly-$\alpha$ absorption due to neutral gas from the companion galaxy enables the detection of Ly-$\alpha$ emission extending from the location of the quasar host to that of the companion galaxy (in both spatial and velocity spaces, panel c in Fig.~1). This implies the presence of gas spread over several tens of kpc, an expected characteristic of merging systems, where material is stretched by tidal forces generated by strong gravitational interactions between the galaxies.

In order to infer the main properties of both galaxies we modelled their spectral energy distribution using constraints on the ultraviolet (UV), optical and mm emission, see Fig.~3. 
We found that the companion and quasar host galaxies are both massive, with stellar masses close to $10^{11} M_{\rm \odot}$, indicative of a major event (as opposed to minor merger when one galaxy is much more massive than the other one). This might have enhanced the star-formation activity in the companion galaxy, located above the main-sequence of galaxies at this redshift\cite{Popesso2023}. The molecular masses derived from the CO emission lines are of the order of $10^{10}$\,$M_\odot$ for both the companion galaxy and that hosting the quasar. We also constrained the mass of the super massive black hole (SMBH) that powers the quasar to be about $10^8 M_{\odot}$ from the emission line width and continuum measurements. Given the $(5-10)\times 10^{46} \rm erg\,s^{-1}$ luminosity, the quasar must be actively accreting matter and radiating at the Eddington limit. The main properties of the galaxies and the quasar are summarized in Table~1.

\renewcommand{\arraystretch}{1.3}
\begin{table}[]
    \centering
    \caption{\textbf{Main properties of the merging galaxies. 
    }
    }
    \begin{tabular}{c c c}
    \hline
    Quantity & Companion galaxy & Quasar host galaxy \\
    \hline
    redshift                    & $2.663\pm0.001$ & $2.656\pm 0.001$ \\
    approaching velocity & \multicolumn{2}{c}{$\sim 550$\,\kms} \\
    projected distance & \multicolumn{2}{c}{$\sim 5$\,kpc} \\
    $\log M_{\rm SMBH}/M_{\odot}$     & -- & $8.3\pm0.4$   \\
    $\log M_{*}/M_{\odot}$  & $10.84^{+0.12}_{-0.12}$ & $10.7^{+0.3}_{-0.3}$  \\
    $\log M_{\rm H_2}/M_{\odot}$ &  $10.00^{+0.05}_{-0.05}$ &  $9.74^{+0.07}_{-0.09}$ \\
    $\log M_{\rm dust}/M_{\odot}$ &  $8.2^{+0.5}_{-0.3}$ & $<8.5$\\
    SFR $[M_{\odot}\,\rm yr^{-1}]$  & $250^{+60}_{-50}$  & $\gtrsim 24^{+17}_{-11}$ \\
    \hline
    \end{tabular}
    \label{tab:mainprop}
\end{table}

\begin{figure*}
\includegraphics[width=1.0\textwidth]{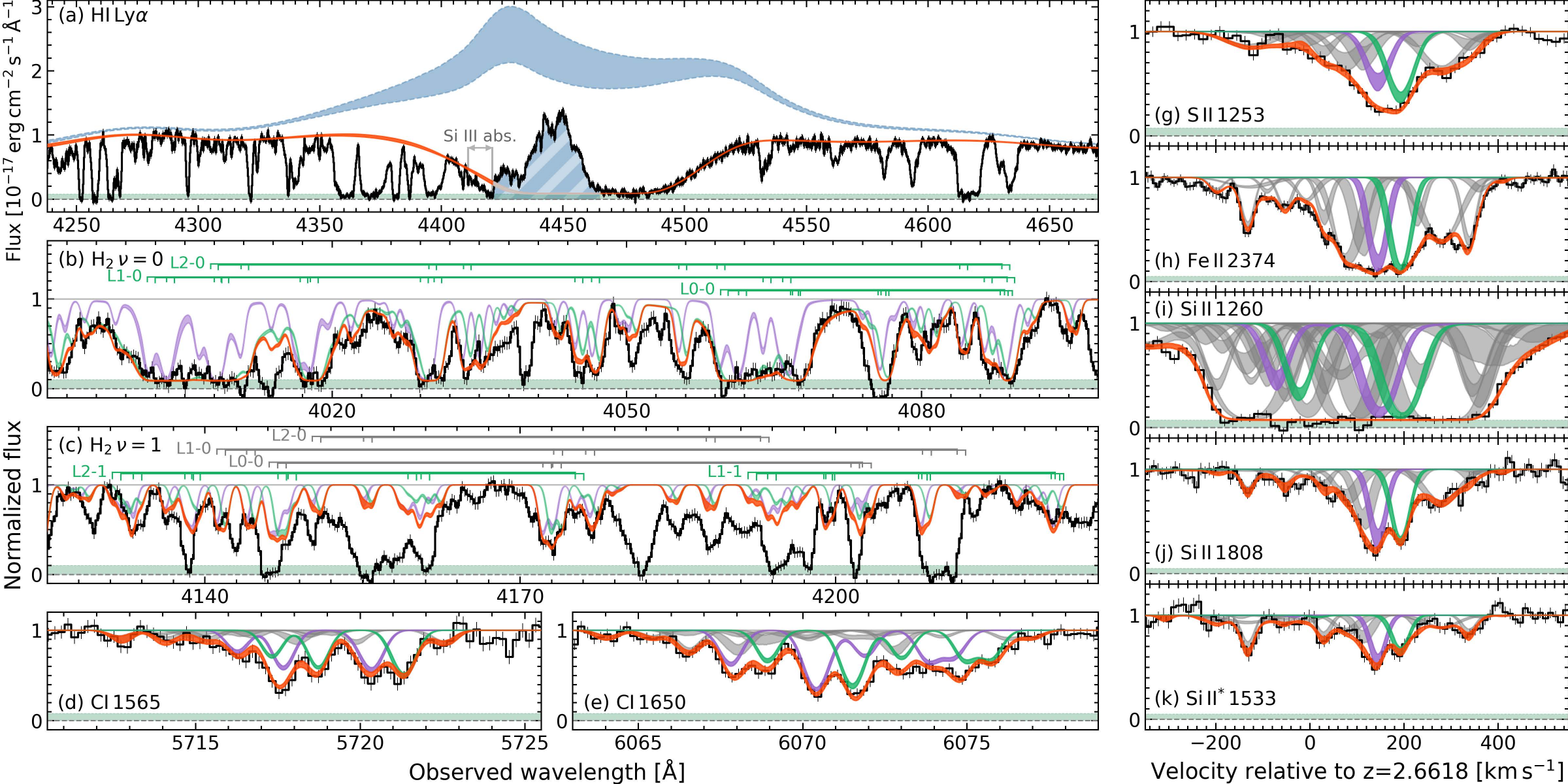}
\caption{
\textbf{Absorption lines in the spectrum of the quasar \qso.} \\
Selected portions of the 1D VLT/X-shooter spectrum of the quasar around absorption lines (marked by text labels) from gas within the galaxy system are shown in black, with error bars representing the $1\sigma$ standard deviation of the flux measurements. The absorption models and their associated uncertainties are colored red, violet, green, and gray (representing total absorption, H$_2$-bearing components, and other components, respectively). The blue shaded region in (\textbf{a}) indicates the reconstructed quasar continuum. The horizontal lines with vertical ticks in \textbf{b} and \textbf{c} indicate the position of individual H$_2$ absorption lines from different vibrational bands marked with text labels on the left. The host galaxy emission is seen as faint residual emission (green) that is not fully absorbed in lines from the merging galaxy system but remains affected by intervening Lyman-$\alpha$ forest lines from the intergalactic medium, which is structured on much larger scales.
\label{fig:xshooter}
}
\end{figure*}

\begin{figure}
\includegraphics[width=1.0\textwidth]{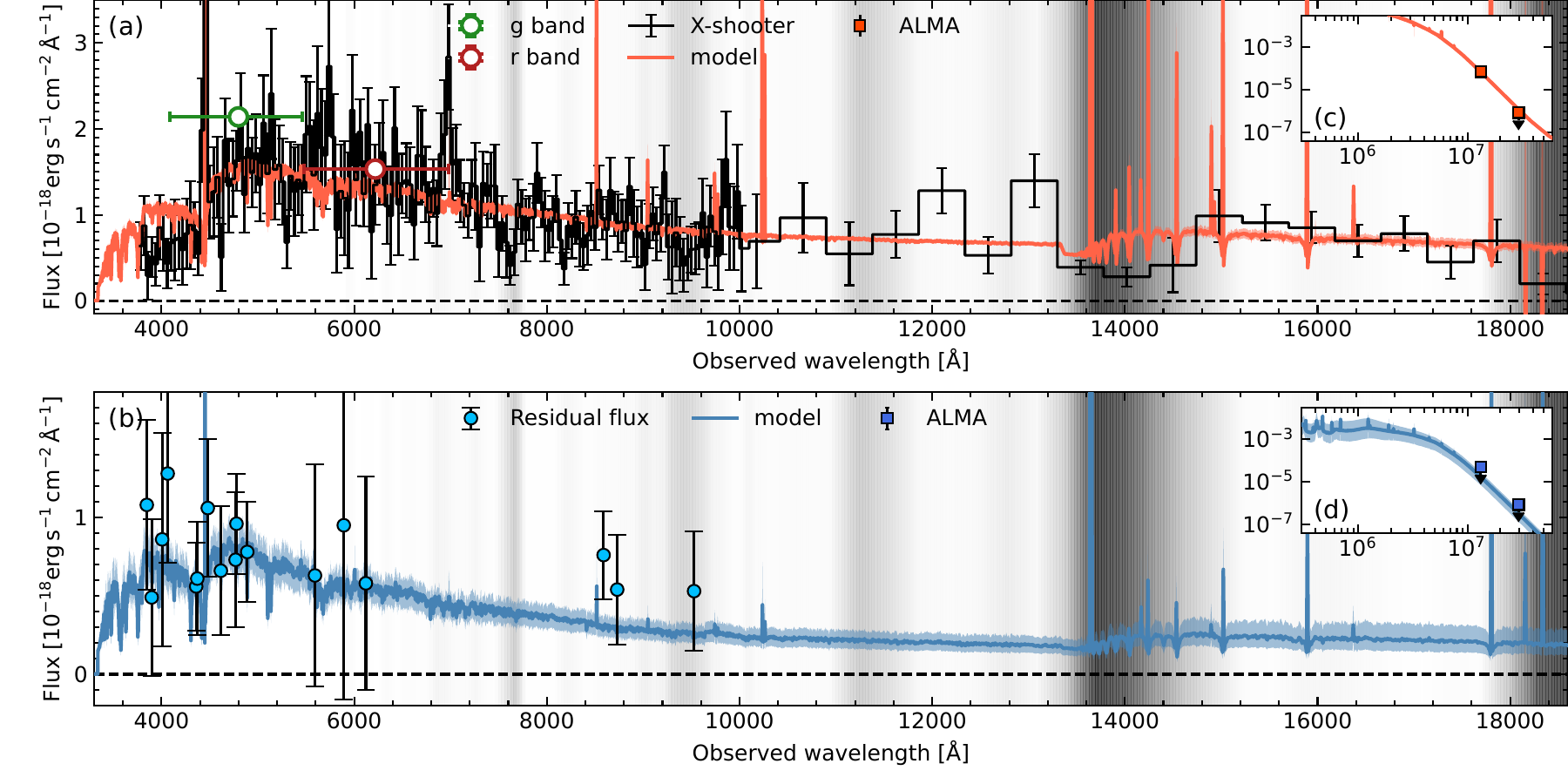}
\caption{
\textbf{Spectral energy distribution of the companion and quasar-host galaxies.}\\ 
\noindent
\textbf{a}, companion. \textbf{b}, quasar-host. The data (black line and colored points) are shown together with the model fit (red and blue shaded regions). The error bars represent 0.683 credible intervals and the standard error of the mean in panels \textbf{a} and \textbf{b}, respectively. The top panel also indicates the HSC photometric data in g and r bands obtained after PSF subtraction of the quasar. The vertical gray stripes indicate the telluric absorptions. \textbf{c, d} The mm-range constrained by ALMA data.
\label{fig:galaxies}
}
\end{figure}

Towards the quasar, we measure the total column density of atomic hydrogen $N(\HI) \approx 10^{21.8}$~\cmsq\ 
(50~M$_{\rm \odot}$\,pc$^{-2}$). 
This high value, which falls in the upper range expected for most high-redshift galaxies (such as those probed by intervening systems) even given the small impact parameter \cite{Rahmati2014, Noterdaeme2014, Krogager2017, Krogager2020}, may arise from material stretched by tidal forces during galaxy interactions.
Moreover, low-ionisation metal absorption lines (see Fig.~2) indicate that the neutral gas is kinematically distributed over $\sim 400\rm\,km\,s^{-1}$ (roughly coinciding with the velocity dispersion of the CO profile, {see panel g of Fig.~1) and has a metallicity which is about one tenth of the Solar value, i.e. consistent with modest chemical enrichment in the companion - as opposed to high enrichment expected in the quasar's host \cite{DiMatteo2004}. The high velocity spread, especially given the metallicity \cite{Ledoux2006}, indicates disturbed kinematics, as expected if arising from a merging system.

The absorbing gas also presents one of the highest H$_2$ column densities ($N(\rm H_2) \approx 10^{21.2}$~\cmsq) ever directly detected in quasar spectra \cite{Balashev2017, Ranjan2018}, distributed across two main velocity components.
But the true exceptional nature of the H$_2$ gas is to be appreciated in its excitation diagram, shown in Fig.~4: the molecular gas presents unusual population of H$_2$ vibrational-rotational levels, with excitation temperatures far exceeding those typically observed in our Galaxy or intervening systems \cite{Balashev2019,Shull2021}. 
Notably, we detect, for the first time in absorption at high redshift, lines from the first excited vibrational state, with energy levels at
$E\gtrsim4000\,\rm cm^{-3}$. 
The excitation observed here is even stronger than what is seen in photo-dissociation regions very close to bright, recently formed stars \cite{Boisse2005}.

Modelling the excitation of H$_2$ with the \texttt{Meudon PDR} code, we infer the UV field to be about a thousand times stronger than the average Milky-Way field. Comparing the quasar luminosity with the derived incident UV flux constrains the absorbing gas to be within at most a few kpc from the quasar (right panel of Fig.~4), similar to the transverse distance between the host and companion galaxy, in line with the merging scenario.

\begin{figure}
\includegraphics[width=1.0\textwidth]{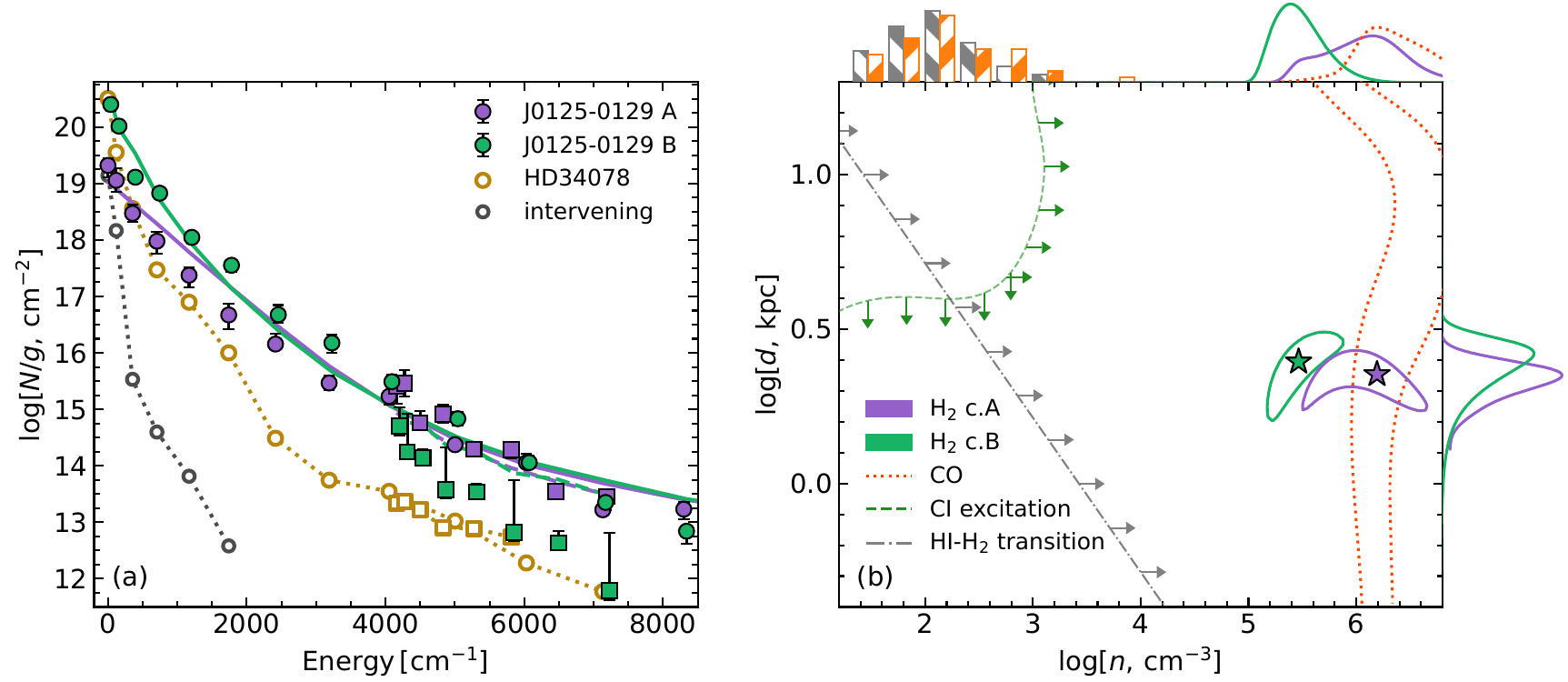}
\caption{\textbf{H$_2$ excitation diagram and physical conditions in H$_2$-bearing medium.}\\ 
\noindent
\textbf{a,}
The population of H$_2$ ro-vibrational levels (circles for the zero vibrational level and squares for the first vibrational level) as a function of the energy of these levels. The values observed towards \qso\ (purple and green for two velocity components) are compared to the excitation seen in intervening systems (grey) \cite{Balashev2019} and to an extreme excitation case in a Galactic system towards a runaway O star (orange) \cite{Boisse2005}. The error bars represent 0.683 credible intervals. The purple and green solid lines represent the best fit models of H$_2$ excitation with \texttt{Meudon PDR} code.
\textbf{b,} 
The constraints on the number density, n, of the absorbing molecular gas and its distance, d, from the quasar, derived from modelling the observed H$_2$ excitation using the \texttt{Meudon PDR} code (green and violet solid lines). The solid contours represent 0.683 credible interval of the 2d posterior distribution. The panel also includes constraints from \CI\ levels (green dashed line), CO abundance (red dotted line), and the HI/H$_2$ transition (gray dashed-dotted line). The 1D marginalized constraints are shown along the top and right axes of \textbf{b}. For comparison, distributions in intervening quasar absorption systems\cite{Balashev2023} and in the Magellanic Clouds\cite{Kosenko2024} are shown as gray and orange histograms, respectively.
\label{fig:H2}
}
\end{figure}

In addition, the presence and the excitation of H$_2$ constrains the gas density to be $n_{\rm H} \sim 10^{5} - 10^{6}$\,cm$^{-3}$. This very high number density is supported by the high excitation of neutral carbon in three fine-structure levels and is further confirmed by the tentative detection of CO absorption, which also requires high density to remain detectable under a strong UV field. The high density and measured column density imply that the molecular clouds are very compact, with sizes, $L\approx N/n_{\rm H}$ (assuming constant density) of less than 0.02~pc. Remarkably, the molecular clouds are then smaller than the expected light beam from the quasar emission line regions, covering only the even much smaller accretion disc. This is independently confirmed by differential dust-reddening between the quasar continuum emission and the emission line regions. The presence of dust around a luminous quasar is also consistent with galaxy interactions\cite{Urrutia2008,Glikman2015}.

The inferred cloud densities are several orders of magnitude higher than those in intervening H$_2$-bearing systems (i.e. far from the quasar), where values $n_{\rm H} \sim 10^{2}$\,cm$^{-3}$ are usually seen. Under normal conditions, cloud densities follow a continuous log-normal distribution\cite{Hennebelle2012}, implying that low-density molecular clouds $\sim 10^2$\,cm$^{-3}$ should outnumber those with $n> 10^{5} $\,cm$^{-3}$ by several orders of magnitude. While we detect high-density molecular clouds, we find no evidence of lower-density ones. This, combined with the high incident UV flux, indicates that the bulk of the diffuse molecular gas exposed to the quasar radiation has been photo-dissociated (within timescales of several years for the constrained UV flux), and only the high density regions survived the intense radiation field.

We further note that while the interception of one molecular cloud by the quasar line of sight is enforced by our selection on the presence of molecular hydrogen absorption lines \cite{Noterdaeme2019}, the detection of a second cloud (we detect H$_2$ in two distinct velocity components separated by $\sim$60\,km\,s$^{-1}$) suggests that the total surface covering factor cannot be extremely small. 
On the other hand, the low average reddening of the background quasar's emission line regions disfavors a high covering fraction of dusty gas over this area. Overall, the volume filling factor of the molecular gas in the region exposed to the quasar radiation should hence be at most of the order of $r/L$, where $r$ is the size of a molecular cloud and $L$ is the kpc-scale path-length across the absorbing galaxy. We infer that the region exposed to the UV radiation from the accretion disk, i.e. a few kpc across at this distance from the quasar, is very clumpy with dense molecular gas surviving in no more than 10$^{-5}$ of the volume. This in line with the growing observational and theoretical evidence that different gas phases in different environments may actually follow a fog-like structure\cite{McCourt2018}, being confined within cloudlets much smaller than the overall extent of the medium\cite{Arav1997}. In our case, the surviving clumps have masses $<0.1 M_{\odot}$ so that they cannot form stars, when the remaining of the companion galaxy, outside the anisotropic quasar radiation field, forms stars at a high pace.

Our findings reveal that we are not only observing an ongoing major merger event, supporting such mechanism as important feeding mechanism for quasars, but also, for the first time, documenting dramatic feedback effects on the internal structure of the gas at unprecedented scales -- 10$^5$ times smaller than directly resolvable through emission studies at high redshifts. Specifically, we have shown that the gas within this system is being transformed into a highly clumpy medium wherever exposed to the intense UV radiation powered by the actively accreting supermassive black hole. This quasar disrupts the molecular gas in a manner analogous to the impact of newborn stars on their parent molecular clouds.


\newpage
\section*{Methods}
\label{sec:methods}
Throughout this work, we use AB magnitudes \cite{Oke1983} and a standard $\Lambda$CDM cosmology with H${_0}=70$ km s$^{-1}$ Mpc$^{-1}$, $\Omega_{\Lambda}=0.7$, and $\Omega_m=0.3$.

\subsection*{Observations} 
We performed UV-to-NIR spectroscopy of \qso\ with X-shooter on the Very Large Telescope at Paranal, obtaining five (three and two) $\approx1$\,hour exposures with two different orientations of position angle PA=90$^{\circ}$ and 0$^{\circ}$, respectively. Observations and data reduction down to the 2D spectrum are described in the survey paper\cite{Noterdaeme2023}. Here, we constructed the skyline solution with a two-degree polynomial fit along the spatial axis, avoiding the quasar and companion galaxy traces ($\sim0.7$ arcsec above the quasar) as well as the edges of the 2D spectrum. For each exposure, we then extracted the one-dimensional quasar spectrum using a Moffat profile, whose parameters (central position and width) were obtained over a large wavelength range while masking cosmic rays and regions containing light from the companion galaxy. Finally, we rescaled and combined all exposures from the different X-shooter arms into a single quasar spectrum.

We obtained ALMA band\,3 and band\,6 data through project 2022.1.01792.S, see Supplementary Table~1.  Typically, 42 - 44 antennas participated in observations, and both the parallel hand correlations were recorded. 
For band\,3, the correlator setup allowed four spectral windows (SPWs) labelled as 23, 25, 27 and 29, centered at 108.794, 94.812, 96.806 and 106.868\,GHz, respectively. The SPW\,23  with a bandwidth of 2\,GHz divided into 128 frequency channels was used to map the radio continuum.  The other three SPWs covering 1.875\,GHz with 480 channels were tuned to search for redshifted CO(3-2), HCO$^+$ and CS spectral lines. 
For band\,6, four spectral windows, i.e., SPW 21, 23, 25 and 27 were centered at 232.652, 234.569, 218.401 and 220.506\,GHz, respectively. The SPWs 21 - 23 and 25 - 27 covered 2\,GHz and 1.875\,GHz split into 128 and 240 frequency channels, respectively. The main objective was to detect redshifted CO(7-6) and \CI(2-1) associated with the quasar in SPW\,27. 
ALMA data were processed using the ALMA pipeline following standard procedures and {\tt robust = 0.5} weighting of visibilities \cite{Hunter2023}. The spectral line cubes properties are summarized in Extended Data Table~1. 

\subsection*{Global properties of the galaxies}

We derived the main global properties of the quasar host and companion galaxies using emission constraints  from our X-shooter and ALMA data, supplemented with Subaru Hyper Suprime Cam (HSC) deep imaging. 

\subsubsection*{Broad-band photometry}
\label{sec:subaru}

We retrieved deep $r$ and $g$ band images of the field from the Hyper Suprime-Cam Legacy Archive 2016 \cite{Tanaka2021} (Extended Data Fig.~1).    
We fitted them with the Galfit package \cite{galfit} using a model consisting of a Point Spread Function (PSF) source (representing the quasar), a sky background, and two additional S\'ersic profiles with index = 1.0, with free position, effective radius, axis ratio, and orientation.
We obtain $g=23.41\pm0.05$ and $r=23.16\pm0.04$ for the main Sérsic source, east of the quasar emission. The second, fainter Sérsic component has $g=24.74\pm0.07$ and $r=26.2\pm0.9$ and extends southwest.  

\subsubsection*{X-shooter spectrum of the companion galaxy}
\label{sec:xshooter}
The companion galaxy's emission is located above the quasar trace in the 2D spectra taken with position angle PA=90$^\circ$. To disentangle this faint emission from that of the quasar, we rebinned each exposure onto a coarser grid corresponding to 200 pixels (40\,\AA) for UVB and VIS arms, and 800 pixels (480\,\AA) for the NIR arm due to sky contamination and weaker fluxes in this region. This was done by subtracting the sky model, summing the 2D spectrum along the spectral axis in each bin, while removing pixels impacted by cosmic rays and strong sky lines.
We then fitted the observed collapsed profiles using two Moffat functions (determined by peak position, width, and normalization), representing the quasar and companion galaxy (Extended Data Fig.~2). 
We followed Bayesian inference by sampling the model parameters using a Python implementation of an affine invariant Monte-Carlo Markov Chain (MCMC) sampler \cite{Goodman2010} within the \texttt{emcee} package \cite{emcee}. For normalization parameters, we used flat priors. For the position and width along the spatial axis, we used Gaussian priors centered at 0'' and the seeing value for the quasar component.
For the companion galaxy component, our priors are based on the results from the broad-band image fitting: the position is taken to be 0.55" from the quasar and the width is taken as the quadratic sum of the unresolved quasar component (i.e. the seeing measured at each wavelength) plus the 0.9"-width derived from Galfit. The derived Moffat profiles were used to compute and correct for slit losses at each wavelength, considering the slit widths of 1.0, 0.9 and 1.2" for UVB, VIS and NIR arms, respectively.

The companion spectrum shown in the top panel of Fig.~3 was then obtained by combining the three individual exposures using the inverse-variance weighted mean. The photometric points obtained using PSF extraction from HSC Subaru imaging are consistent with this spectrum.  

\subsubsection*{The quasar host galaxy as residual emission in saturated absorption lines}
\label{sec:host}

We constrained the quasar host galaxy emission from residual light in the dark troughs of heavily saturated absorption lines (Ly$\alpha$, \SiII\,1260\,\AA, \OI\,1302\,\AA, \CII\,1334\,\AA, \AlII\,1260\,\AA\ and series of H$_2$ and \FeII\ lines) in the quasar spectrum. The 2D spectrum (Extended Data Fig.~3) shows that this residual emission is aligned with the unresolved quasar trace and remains similar regardless of the position angle, indicating that it comes mainly from the quasar host galaxy, covered by both slit positions, not from the companion galaxy (covered by the slit at PA=90$^\circ$ but not at PA=0$^\circ$).
Residual emission is also seen in the blue part of the spectrum, as a non-zero level in saturated H$_2$ absorption lines. However, this emission is absent in the troughs of strong Ly$\alpha$ forest lines, which fully absorb light from the background objects (quasar, host and companion galaxies) due to their origin in large-scale intergalactic structures. This confirms that the residual flux in the $z=2.66$ lines is real and not a data reduction artifact. The constraints on the quasar host galaxy spectrum (bottom panel of Fig.~3) were derived by combining all five exposures, each extracted using a Moffat profile along the quasar trace to minimize any contribution from the companion galaxy.

\subsubsection*{Millimetre line and continuum emission}

 The ALMA cubes were blindly searched for line emission using the \HI\ Source Finding Application \texttt{SOFIA v2.0} \cite{Serra2015, Westmeier2021}. We set \texttt{SOFIA} to subtract the residual continuum and use the smooth+clip (S+C) algorithm with a combination of spatial and spectral kernels matched to the expected signal size in a given band. We set the threshold to 3.5$\sigma$ for detection. For the moment 1 and 2 maps, we used only those pixels where the signal was detected at 1.5 times the local rms value. We detect CO(3-2), CO(7-6) and even \CI(2-1) line emission from the companion galaxy (see Fig.~1 and Extended Data Fig.~4). The quasar is detected only in CO(7-6) line emission (shown in Fig.~1). The integrated line fluxes estimates (including 3$\sigma$ upper limits for nondetections) from the moment-0 maps are provided in Extended Data Table~1. For CO line ratio map of the companion galaxy (shown in Extended Data Fig.~4), we smoothed the CO(7-6) map to the lower resolution of CO(3-2) map. 

We also examined continuum images for radio emission associated with the galaxy and the quasar. In band\,3, no continuum emission is detected in the multiband radio continuum image at 101.9\,GHz made using all four SPWs with {\tt robust = 0.5} weighting (see panel (e) in Extended Data Fig.~4). The synthesized beam and the rms noise are $0.177^{\prime\prime}\times0.141^{\prime\prime}$ and 8.3\,$\mu$Jy\,beam$^{-1}$, respectively. We adopt a $3\sigma$ upper limit of 25\,$\mu$Jy for the quasar and the companion galaxy. The band\,6 multiband image at 226.5\,GHz with an rms of 14.8\,$\mu$Jy\,beam$^{-1}$ exhibits radio emission (panel (d) in Extended Data Fig.~4) with  
flux densities of 420\,$\mu$Jy and 220\,$\mu$Jy at the positions of the galaxy and the quasar, respectively. However, the corresponding emission is not consistently detected in images from individual SPWs with rms noise of $\sim$25\,$\mu$\,Jy\,beam$^{-1}$. Based on measurements in SPWs 21, 23 and 25 of band 6, we adopt continuum flux densities at $\approx226$\,GHz to be  330$\pm$30\,$\mu$Jy and $<$180\,uJy for companion and host galaxies, respectively.

\subsubsection*{Spectrophotometric modelling of the galaxies} 
\label{sec:sed_fit}
We used the \texttt{bagpipes} code \cite{Carnall2018} to fit the photometry data from Subaru, the ALMA constraints on the mm-emission continuum, and the spectral energy distribution (SED) obtained from X-shooter for both the companion and host galaxies. We assumed a double power law star formation history (SFH) model\cite{Carnall2019} and standard dust extinction law\cite{Calzetti2000}. 
We treated the metallicity as a free parameter, recognizing that it could differ between the central regions of the companion galaxy and that along the quasar line of sight where we had the prior from absorption lines. We assumed a stellar birth cloud lifetime of 10 Myr and a velocity dispersion of 500\,\kms. While we considered nebular emission to fit emission lines (Ly$\alpha$ and {\sc C\,III}] lines are prominent in the companion galaxy spectrum), we acknowledge that these could be impacted by the large characteristic bin size and potential quasar contribution. For dust emission from the companion galaxy, given the likely subsolar metallicity, we allowed variations in $U_{\rm min}$ (the lower cutoff of the starlight intensity distribution) and $\gamma$ (the fraction of dust heated by starlight)\cite{Draine2007}. We used Gaussian priors on the redshift and metallicity from the CO(7-6) emission and the absorption line analysis, respectively. Uniform priors were applied to the star-formation law slope (in decimal scale) and in log-scale for other parameters. The parameters and their constraints, derived from the posterior probability function using nested sampling within the {\sc bagpipes} code, are presented in Extended Data Table~2. While some parameters are not well constrained, due to limited observational data, others, like stellar mass and extinction, are precisely determined.

\subsubsection*{Mass estimates}
\label{sec:masses}
\bmhead{ Molecular gas}
\noindent The observed CO line fluxes correspond to luminosities\cite{Solomon2005} $L^{'}_{\rm CO}$= ($2.5\pm0.3$), ($9.7\pm1.2$), and ($1.4\pm0.3$) $\times 10^9$ K\,km\,s$^{-1}$\,pc$^2$ for CO(7-6) and CO(3-2) lines in the companion galaxy, and CO(7-6) line in the quasar host galaxy, respectively. These are converted into molecular masses using $M_{\rm H_2} = \alpha L^{'}_{\rm CO}(J-(J-1))/r_{J1}$, where $\alpha$ is the CO(1-0) to H$_2$ conversion factor and $r_{J1}$ is the (J-(J-1)) to (1-0) CO line ratio. Assuming \cite{Boogaard2020} $r_{31}=0.8$ and $r_{71}=0.2$ we obtain $M_{\rm H_2} = (1.0\pm0.1)\times10^{10}\,M_{\odot}$ (companion) and $M_{\rm H_2} = (5.5\pm0.1)\times 10^{9}\,M_{\odot}$ (host), adopting $\alpha=0.8\,M_{\odot} (\rm K\,km\,s^{-1}\,pc^{2})^{-1}$ as suggested by measurements in AGN, mergers and starburst environments\cite{Carilli2013,Bolatto2013,Sargent2014,CalistroRivera2018}.\\

\bmhead{Dust}
The dust masses are estimated from the observed ALMA band 6 fluxes, assuming a gray body spectrum with dust opacity $\propto\nu^{\beta}$ and $\beta=0.7$ (consistent with the relative fluxes of ALMA bands 3 and 6), and a dust temperature range of 20 to 60\,K, yielding $(1-5)\times10^8$\,$M_{\odot}$ for the companion and $<3\times10^8$\,$M_{\odot}$ for the host galaxy, respectively.\\

\bmhead{SMBH}
The mass of the SMBH powering \qso\ is estimated from the widths of the \CIV\ (FWHM=$3830\pm30$\,\kms) and H$\beta$ (FWHM=$2730\pm30$\,\kms) emission lines together with continuum measurements. Correcting for dust reddening, we measure fluxes of $2.5$ and $0.3\times 10^{-17}\,\rm erg\,cm^{-2}\,s^{-1}\,$\AA$^{-1}$ at 1350 and 5100~\AA, respectively, translating to $\log \lambda L_{\lambda}/ (\rm erg\,s^{-1})= 45.9$ and $45.5$. This yields $\log\,M_{\rm SMBH} / M_{\odot} = 8.3\pm0.4$ (from \CIV) and 8.0$\pm$0.4 (from H$\beta$) using standard scaling relations\cite{Vestergaard2006}. These values align with the $M_{\rm SMBH}$-$\sigma_*$ relation \cite{Kormendy2013}, where the buldge velocity dispersion $\sigma_* \simeq 200$~\kms, is estimated from the stellar mass $M_* \simeq 5\times 10^{10}\,M_{\odot}$ obtained above \cite{Zahid2016}.
The estimated SMBH mass corresponds to an Eddington luminosity $L_{\rm edd}$ of a few times $10^{46} \rm erg\,s^{-1}$, close to the bolometric luminosity of the quasar, obtained from the flux at  1450\,\AA\, and scaling relations\cite{Runnoe2012}, indicating that the SMBH is accreting close to its Eddington limit. 

\subsection*{Absorbing gas analysis \label{sec:absorption}}

\subsubsection*{Differential dust extinction} 
\label{sec:dust}

The spectrum of \qso\ appears clearly reddened compared to the X-shooter composite quasar spectrum\cite{Selsing2016}, as shown in Extended Data Fig.~5. Following the quasar pair method\cite{Balashev2019,Wang2004,Srianand2008, Zhang2015, Noterdaeme2017}, the best matching is obtained by reddening the composite with an Small Magellanic Cloud (SMC) like extinction curve\cite{Gordon2003} and visual extinction $A_{\rm V}\approx0.20\pm0.10$. However, this significantly underpredicts the emission lines, {Si\,{\sc IV}}, {\sc C\,III}] and {\sc C\,IV}, suggesting a differential reddening between the continuum and the emission lines. Combining the reddened continuum with unreddened emission lines from the composite provides a much better match, indicating that dust is mostly confined in small clouds that cover the accretion disc but not the larger emission line regions. We further note that while the heights of the emission lines are now well reproduced, the observed lines are narrower than in the composite (see inlay panels in Extended Data Fig.~5). This is again consistent with the differential reddening, as the wings of the emission lines (i.e. the broad emission component) are produced in a small region with high kinematics closer to the accretion disc, while the core of the lines (the narrow emission) are produced over larger areas further from the central engine. Similar patchy dust obscuration has also been invoked to explain the unusual line shapes observed in extremely red quasars \cite{Hamann2017} and in some AGNs in the local Universe \cite{Veilleux2013}.

\subsubsection*{Line profile fitting method}
We modelled the absorption lines using multi-component Voigt profile fitting using \texttt{spectro} package. The model parameters were constrained following a Bayesian approach to derive the posterior distribution function using the affine invariant MCMC sampler \cite{Goodman2010}.
We report constraints on the model parameters using maximum a posteriori probability and 0.683 highest posterior density estimates of the 1D marginalized probability distribution function. Priors were set using uniform distributions for parameters such as redshifts ($z$), Doppler parameters ($b$), and logarithm of the column densities ($\log N$), unless otherwise specified. The model was compared to data via the likelihood function, focusing on visually selected spectral regions that are free from blends, cosmic-ray or skyline residuals. We assumed Gaussian-distributed uncertainties around central pixel values, with dispersion given by the flux uncertainty.

Remarkably, most saturated absorption lines at $z=2.66$ do not reach the zero level, indicating partial coverage of the emission source by the absorbing medium. A partial coverage factor, $C_f$, was then included in the model\cite{Bergeron2017}. To reduce the number of free parameters, a single $C_f$ value was assumed for all components at all wavelengths, as supported by the nearly constant residual emission across wavelengths. 

\subsubsection*{Neutral hydrogen \label{sec:HI}}
The total \HI\ column density is determined by fitting the Ly-$\alpha$ absorption profile together with the unabsorbed continuum, initially estimated using the reddened quasar composite and adjusted with an 11th order Chebyshev polynomial. The pixels corresponding to residual Ly$\alpha$ emission (mainly within 4430–4460\,\AA) were excluded from the fit. We obtained $\log N(\mbox{HI}) = 21.911^{+0.016}_{-0.014}$ at $z= 2.66347^{+0.00024}_{-0.00027}$ and a zero-level flux level $7.8^{+0.5}_{-0.2} \times 10^{19}$ erg\,cm$^{-2}$\,s$^{-1}$\,\AA$^{-1}$, seen as residual emission at $\sim4460-4490$\,\AA\ in Extended Data Fig.~3.

\subsubsection*{Molecular hydrogen \label{sec:H2}}
 
The unabsorbed continuum over H$_2$ lines was re-constructed by eye using B-spline interpolation and guided by the reddened composite quasar spectrum. A two-component model (components $A$ and $B$) was necessary to fit the absorption lines (see lines around $4084$\,\AA\ and $3930$\,\AA). We tied Doppler parameters ($b$) for $J=0$ and $1$ rotational levels, assuming that they trace the bulk of the cloud and are co-spatial. Higher rotational levels were allowed different $b$-values, constrained to increase with rotational level \cite{Lacour2005, Noterdaeme2007, Balashev2009} using a penalty function added to the likelihood\cite{Noterdaeme2021, Kosenko2021}. We also penalized non-physical models where $T_{j,j+2} > T_{j+2,j+4}$, considering the ortho- and para- H$_2$ independently. The fit to H$_2$ is shown in Extended Data Fig.~6 and constraints on the model parameters are summarized in Supplementary Table~2. We found the the total column density is $\log N({\rm H}_2) = 21.19^{+0.02}_{-0.01}$. The levels $J=0$ and 1 provide estimates of the kinetic temperatures $T_k\approx T_{01}=320^{+40}_{-20}$~K and $300^{+20}_{-20}$ K for components $A$ and $B$, respectively.

\subsubsection*{Metal lines}
\label{sec:metal}
The absorption profile of metal lines extends over $\sim 400$ km\,s$^{-1}$ (see Fig.~2)
We fitted the data using an 18-component model, with Doppler parameters tied between different low-ionization metals for each component, i.e. implicitly assuming turbulence-dominated broadening.
The unabsorbed continuum was locally reconstructed over each absorption line through B-spline interpolation using the surrounding unabsorbed spectral regions. We used Gaussian prior on the Doppler parameter with mean $6\,\rm km\,s^{-1}$ and dispersion $2\,\rm km\,s^{-1}$, in agreement with typical values measured in high-resolution studies. The fit results are reported in Supplementary Table~3 and 4. We obtain the average metallicity from the ratio of the total column density of the volatile element zinc to that of the total hydrogen column density $N_{\rm tot}(\rm H) = N(\HI) + 2 N(\rm H_2)$, giving $\log Z/Z_{\rm \odot}= - 0.92\pm0.03$ (respective to Solar\cite{Asplund2009}). 

\subsubsection*{Neutral carbon}
\label{sec:CI}
We fitted the \CI\ absorption lines from the fine structure levels of the ground electronic state using a six-component model, identifying two prominent major components aligning with the H$_2$-bearing components. This association is consistent with findings in typical H$_2$-bearing DLAs \cite{Noterdaeme2018}. The detection of \CI\ in components without H$_2$ also indicates the presence of the relatively dense neutral gas\cite{Balashev2024}. We tied the Doppler parameters of the three fine-structure levels for each component, as these are typically co-spatial within the cloud. Consequently, we also tied the column densities of the \CI\ levels using a one-zone model, assuming radiative pumping\cite{Balashev2017} and collisional excitation by H$_2$ or \HI, and Helium\cite{Schroder1991, Abrahamsson2007, Staemmler1991}. We used temperature priors from H$_2$ whenever detected and a flat prior otherwise. We accounted for excitation by the Cosmic Microwave Background, using the expected temperature $T_{\rm CMB}(z) = (1+z) \times 2.725 \approx 10$\,K at $z=2.662$. We used Gaussian priors with a mean of $6\,\rm km\,s^{-1}$ and dispersion of $2\,\rm km\,s^{-1}$ on Doppler parameters. The fit to \CI\ fine-structure levels is shown in Extended Data Fig.~7 with interval estimates on the model parameters in Supplementary Table~5. 

\subsubsection*{CO absorption}
We tentatively detect CO absorption in $\rm A-X$ bands (1-0), (3-0), (4-0), (6-0) and (7-0) at $z=2.66$. 
Since lines from individual rotational levels within each band are not resolved, we fitted the band profiles assuming the rotational population follows a Boltzman distribution, setting the prior on the kinetic temperature to be $\log T[\rm K]=2\pm0.3$, in between the temperature obtained from the ortho-para ratio of H$_2$ and the typical values in CO-bearing medium. We used a two-component model with priors on the Doppler parameters $b=3\pm1$\,km\,s$^{-1}$ and redshifts fixed to those from H$_2$ absorption lines, obtaining $\log N(\rm CO) < 13.6$ and $13.99^{+0.15}_{-0.09}$ in components A and B, respectively. In Extended Data Fig.~8 we compare the stack of observed CO bands to that of the synthetic profiles\cite{Noterdaeme2018}.

\subsection*{Physical conditions in the absorbing gas}

\bmhead{Constraints from H$_2$ excitation}
\label{sec:H2_exc}
We used the \texttt{Meudon PDR} code (version 1.5.4) \cite{LePetit2006} to model the column densities of H$_2$ in various rotational-vibrational levels. We considered a constant-density medium with 0.1 Solar metallicity, exposed to UV radiation from the quasar on one side\cite{Noterdaeme2021}. 
Other parameters were: cosmic ray ionization rate $10^{-16}$ s$^{-1}$ per H$_2$ molecule, LMC extinction curve with $R_V=2.6$, standart MRN dust size distribution with dust-to-gas mass ratio of 0.01 and turbulent velocity of 2\,\kms. We note that these additional parameters have little effect on the H$_2$ excitation and thermal balance in the UV-dominated regime. The default Mathis field was scaled to mimic the quasar UV radiation at a given distance $d$. The choice of the overall shape of the radiation field does not affect the results since photoexcitation and photodissociation of H$_2$ involve the wavelength region $\sim1100-910$\,\AA, where both Mathis and typical AGN fields are approximately flat. 
We constructed a two-dimensional grid of models in the number density–distance ($n-d$) plane, trimming each model to match the observed total H$_2$ column density\cite{Klimenko2020,Kosenko2024}. We computed each model's likelihood using the observed column densities in the various excitation levels. The resulting parameter estimates are shown in Fig.~4. These were obtained using Bayesian approach assuming derived likelihood and flat priors on $d$ and $n$ in log space to emulate a wide distribution.

\bmhead{Constraints from CO/H$_2$}
We obtained constraints on the physical conditions from the total observed CO and H$_2$ column densities in component $B$ using the same grid of models. Taken as face value, the presence of CO suggests slightly larger number densities than derived from the excitation of H$_2$ in component $B$. This could result from the cloud having higher densities in the central regions, where CO is more easily formed. 

\bmhead{Constraints from \CI\ fine structure}
The \CI\ line profile fit directly constrains the physical conditions, as we used an excitation model to link the population of the fine-structure levels. Because the excitation of the fine-structure levels are close to the ratio of statistical weights, \CI\ provides lower and upper limits on the number density and distance to the AGN, respectively, as highlighted in  Fig.~4.

\bmhead{Constraints from \HI/H$_2$}
Based on the analytical theory of the \HI/H$_2$ transition\cite{Sternberg2014, Bialy2016}, we can only obtain a limit in the $n-\rm UV$ \cite{Noterdaeme2019}, as it is not feasible to determine the fraction of the total \HI\ column density associated with the H$_2$ components. Nonetheless, this limit agrees with the other constraints.

\bmhead{Data Availability}
The Subaru imaging data are available on the \texttt{Hyper Suprime-Cam Legacy Archive} (\texttt{HSCLA}; \url{https://hscla.mtk.nao.ac.jp/}). VLT spectroscopic data were collected at the European Southern Observatory under ESO programme 105.203L (PI: P. Noterdaeme) and publicly available through the ESO science archive at \url{https://archive.eso.org/cms.html}. ALMA data collected under program ID 2022.1.01792.S (PI: S. L{\'o}pez) are publicly available at ALMA science archive at \url{https://almascience.eso.org/aq/}.

\bmhead{Code Availability} 
Astropy \cite{astropy:2022},
Bagpipes \cite{Carnall2018}, 
Galfit \cite{galfit},
Matplotlib \cite{Matplotlib2007}, 
NumPy \cite{Numpy2020},
Photutils \cite{photutils},
SciPy \cite{Scipy2020},
SOFIA \cite{Serra2015, Westmeier2021},
Emcee \cite{emcee},
\texttt{spectro} \url{https://github.com/balashev/spectro}
\texttt{Meudon PDR} \url{https://pdr.obspm.fr/}



\bmhead{Acknowledgments}
SB is supported by RSF grant 23-12-00166 and thanks IAP for hospitality where part of this work was done. The research leading to these results received support from the French {\sl Agence Nationale de la Recherche} under ANR grant 17-CE31-0011-01/project “HIH2” (PI Noterdaeme). NG acknowledges NRAO for generous financial support for the sabbatical visit at Socorro during which a part of this work was done.

\bmhead{Author Contributions}
S.A.B. and P.N. led the analysis and wrote the manuscript. S.A.B. modeled the physical conditions. P.N. designed the optical observations. N.G. performed the analysis of the ALMA data. J.-K.K. reduced the X-shooter data, with post-processing by R.C. F.C., S.L. and A.O. contributed to the ALMA observing proposal. P.P. and R.S. contributed to the VLT proposal. All authors contributed to the text and aided the interpretation of the results.

\bmhead{Author Information}
Correspondence and requests for materials should be addressed to Sergei Balashev. The authors declare no competing interests.

\newpage

\section*{Extended Data}

\renewcommand{\figurename}{\textbf{Extended Data Fig.}}
\renewcommand{\tablename}{\textbf{Extended Data Table}}
\setcounter{figure}{0}  
\setcounter{table}{0}  

\begin{figure*}[!h]
\centering
\includegraphics[trim={0.0cm 0.0cm 0.0cm 0.0cm},clip,width=\textwidth]{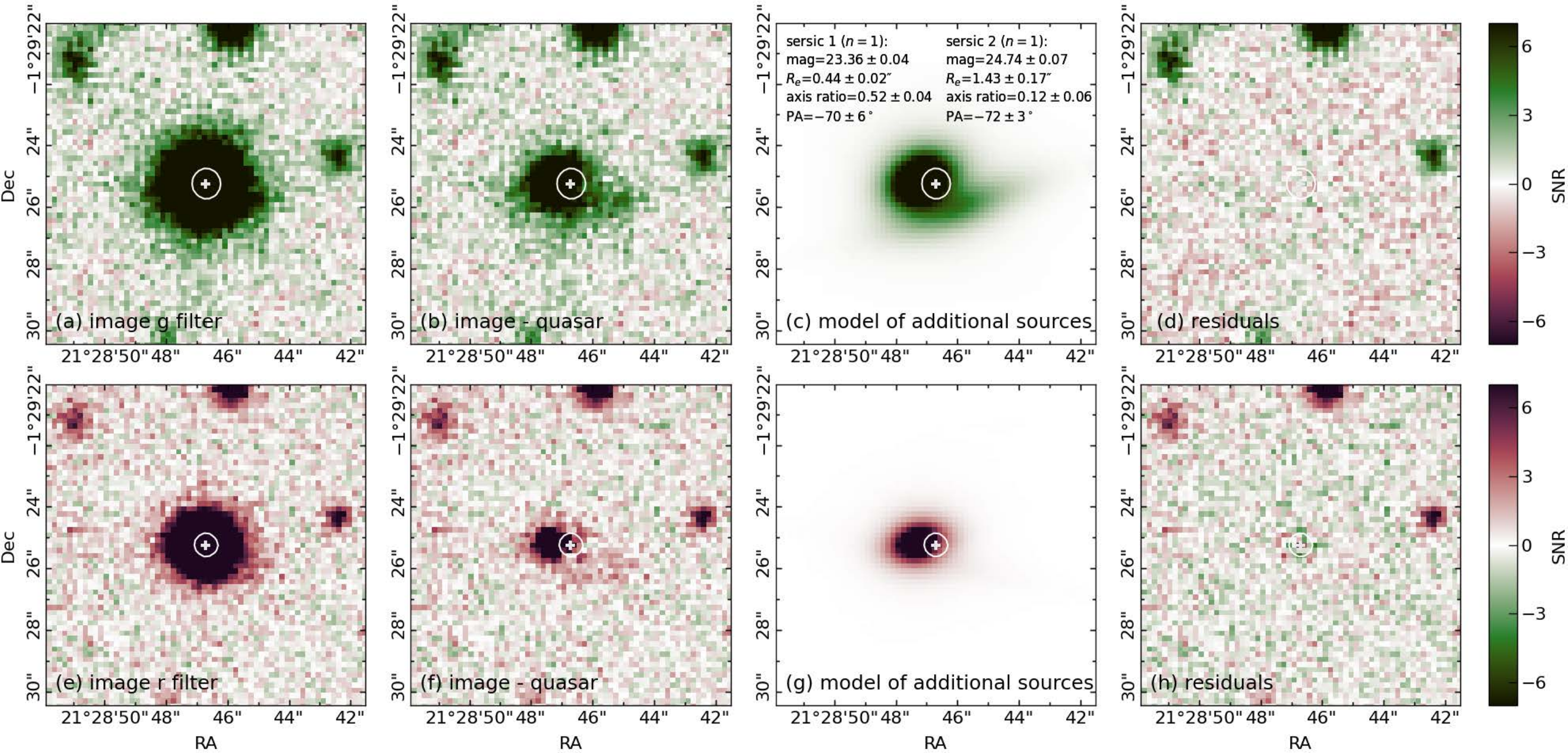}
\caption{
\textbf{Hyper Suprime-Cam deep images of the field around \qso.} \textbf{a, e}, the original image. 
\noindent
\textbf{b, f}, quasar's point spread function (PSF) subtracted image. \textbf{c, g}, model of additional sources. \textbf{d, h}, residuals. The top and bottom panels correspond to $g$ and $r$ filters, respectively. The blue circle in each panel denotes the full width at half maximum (FWHM) of the PSF. The white cross and circle indicate the position of the quasar and the FWHM of the point spread function, respectively.
\label{fig:subaru}
}
\end{figure*}

\setlength{\tabcolsep}{2pt}
\begin{table}
\caption{{\bf Details of ALMA spectral line cubes and detected line emission.} 
\label{tab:almaline}
}
\vspace{-0.4cm}
\begin{tabular}{ccccccccc}
\hline
\hline
Band & SPW & Target line & $\nu_{rest}$  & Beam$^{a}$ & Res.$^b$ & rms$^{c}$   & \multicolumn{2}{c}{Line flux$^{d}$}  \\
     &     &             &     [GHz]     &            &   [\kms] &  [mJy$\times$ & \multicolumn{2}{c}{[Jy$\,$\kms]}  \\
     \multicolumn{6}{c}{} & \,beam$^{-1}$] & Companion & Quasar \\
 \hline          
 3 &  25  & CO(3-2)   & 345.7960  & $0.210^{\prime\prime}\times0.166^{\prime\prime}$, $-51.5^\circ$  & 12.4 & 0.29 & 0.271 $\pm$ 0.034 & $<$0.10 \\ 
 '' &  27  & HCO+(4-3) & 356.7343  & $0.207^{\prime\prime}\times0.161^{\prime\prime}$, $-51.1^\circ$ & 12.0 & 0.29 & $<$0.10      & $<$0.10 \\
 '' &  29  & CS(8-7)   & 391.8468  & $0.190^{\prime\prime}\times0.150^{\prime\prime}$, $-52.0^\circ$ & 10.9 & 0.32 & $<$0.10      & $<$0.10 \\
 6  &  27  & CO(7-6)   & 806.6518  & $0.121^{\prime\prime}\times0.101^{\prime\prime}$, $+72.4^\circ$ & 10.6 & 0.37 & 0.383 $\pm$ 0.044 & 0.207 $\pm$ 0.039 \\ 
 '' &  ''  & \CI(2-1)   & 809.3420  &  ''   &   '' &  ''  & 0.170 $\pm$ 0.036 & $<$0.12 \\ 
\hline
\hline
\end{tabular}
$^a$ Spatial resolution i.e., the synthesized beam in terms of major and minor axis, and position angle. $^{b}$ Spectral resolution in the quasar frame. $^{c}$ Spectral rms. $^{d}$ For non-detections 3$\sigma$ upper limits estimated assuming a line FWHM of 100\,\kms.
\end{table}

\begin{figure*}
\centering
\begin{tabular}{cccc}
\includegraphics[trim={0.0cm 0.0cm 0.0cm 0.0cm},clip,width=0.33\textwidth]{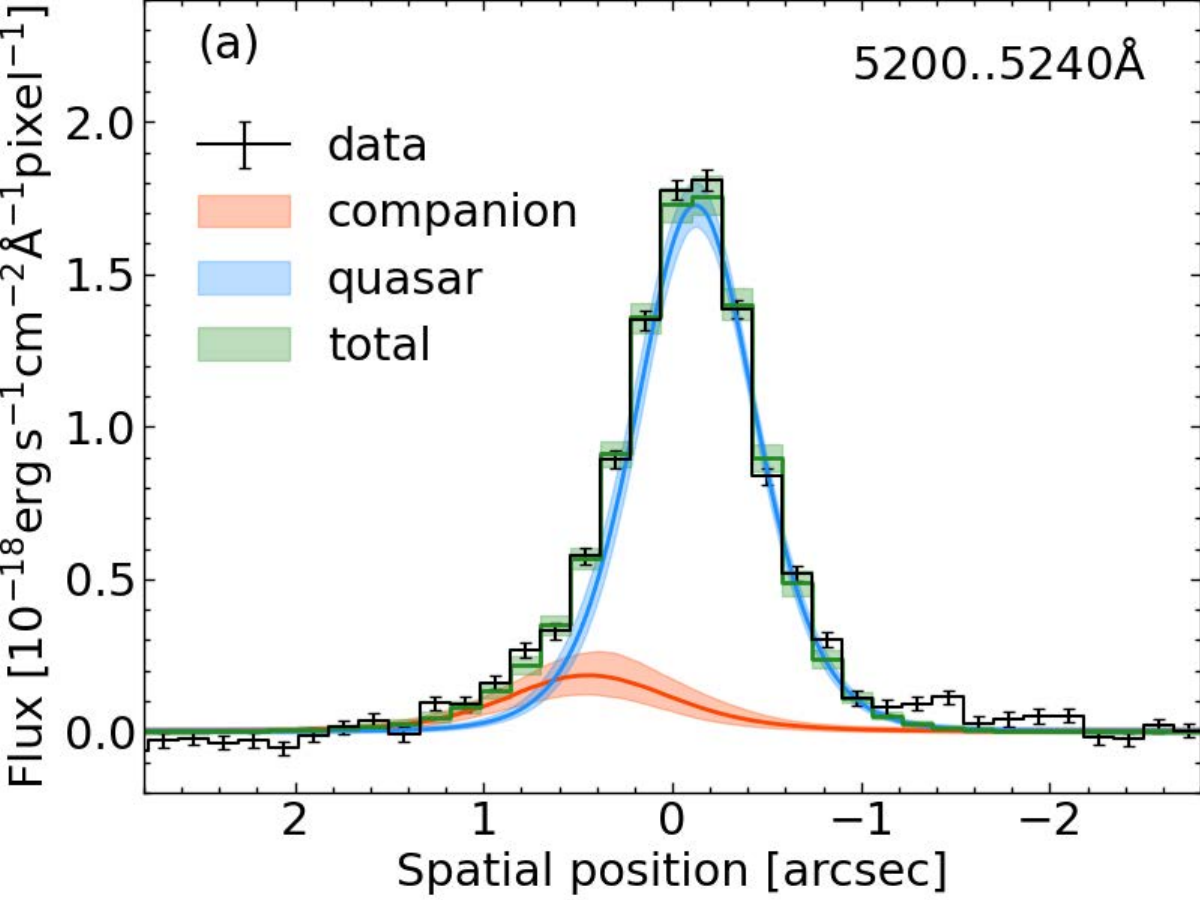} &
\includegraphics[trim={1.2cm 0.0cm 0.0cm 0.0cm},clip,width=0.31\textwidth]{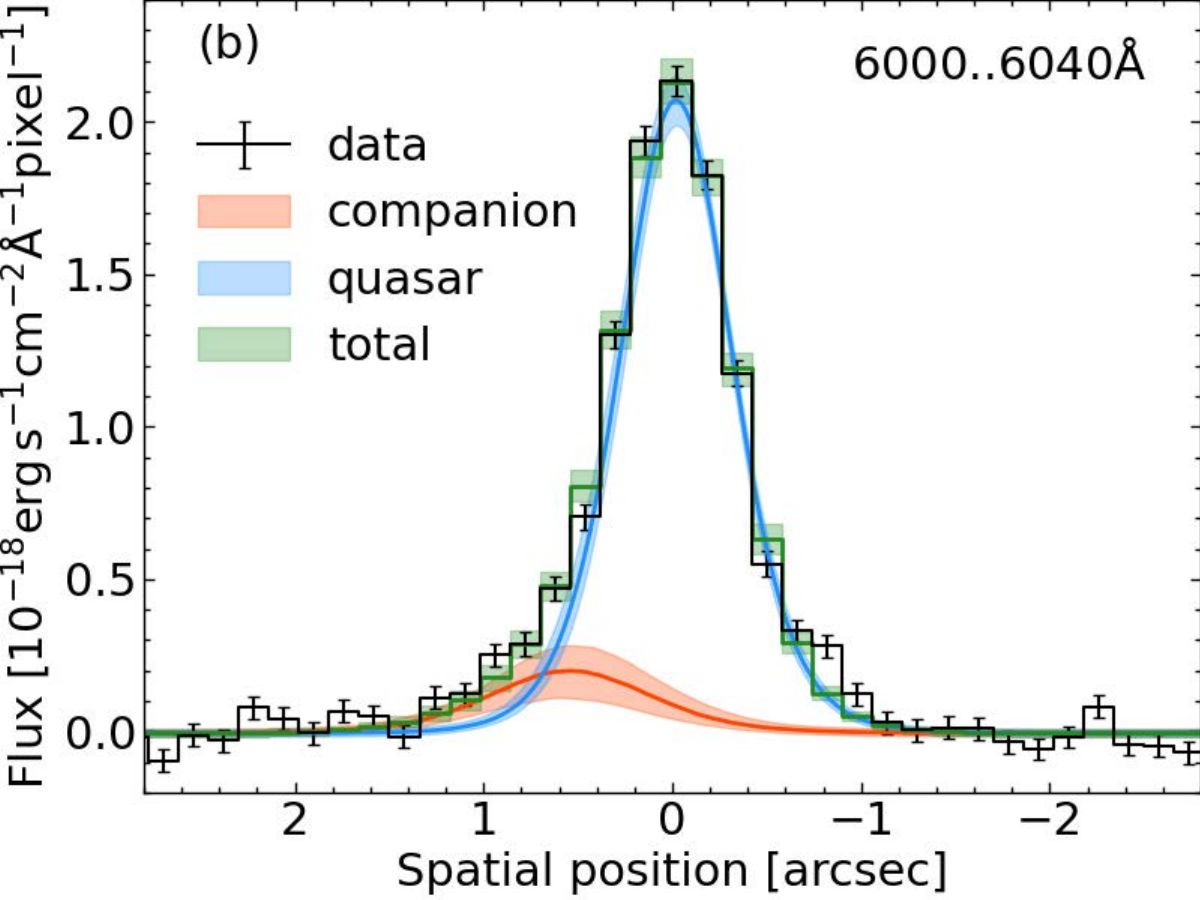} &
\includegraphics[trim={1.2cm 0.0cm 0.0cm 0.0cm},clip,width=0.31\textwidth]{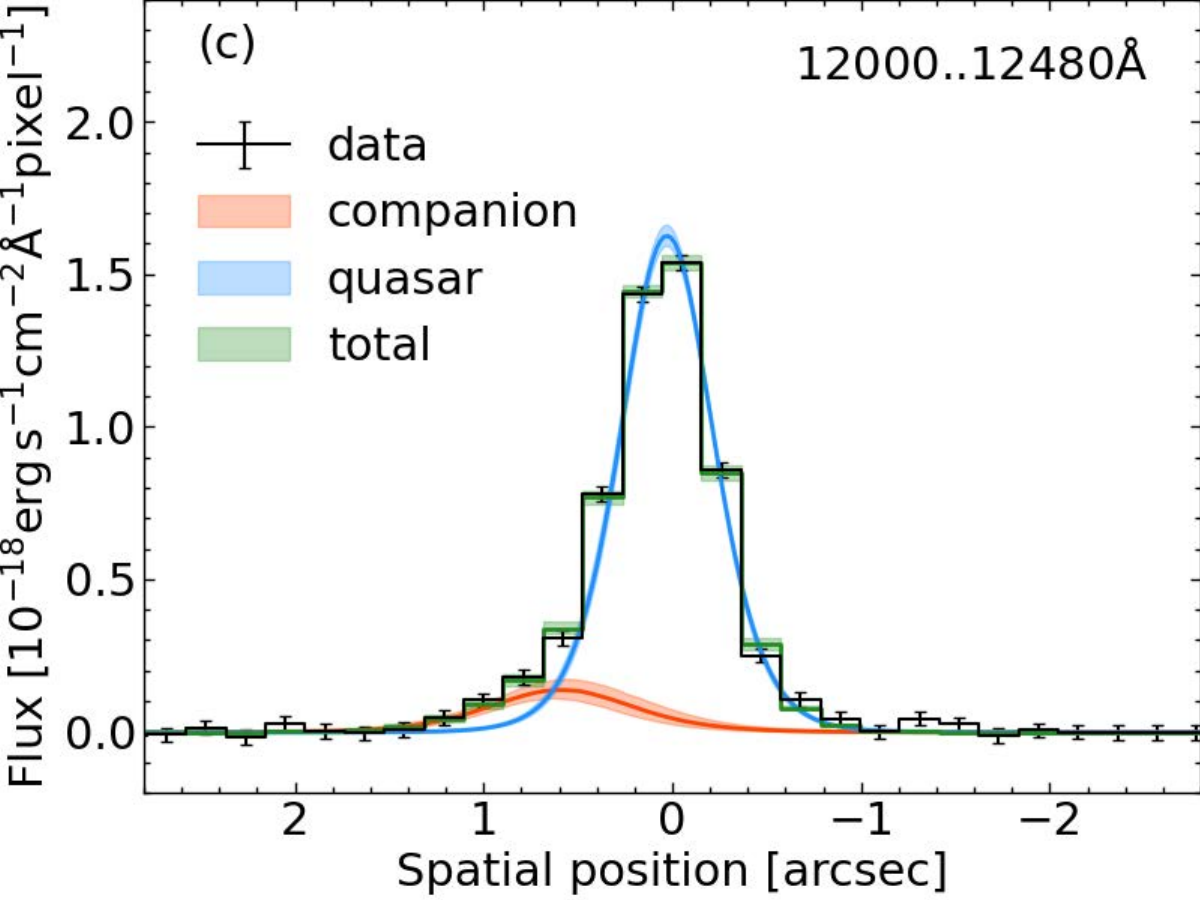} \\
\end{tabular}
\caption{
\textbf{Spatial profile of the 2D X-shooter spectrum of \qso.} 
\noindent
The data (black lines with error bars) is shown here collapsed into three \textbf{a, b, c}, wavelength regions (indicated in the top right corner). The spatial profile is modeled by the sum (green) of two Moffat profiles, corresponding to the quasar (blue) and the companion galaxy (red). The error bars indicate the standard deviation of the collapsed flux, while the colored stripes represent the 0.683 credible interval of the models.
\label{fig:comp_extraction}
}
\end{figure*}

\begin{figure*}
\centering
\includegraphics[trim={0.0cm 0.0cm 0.0cm 0.0cm},clip,width=\textwidth]{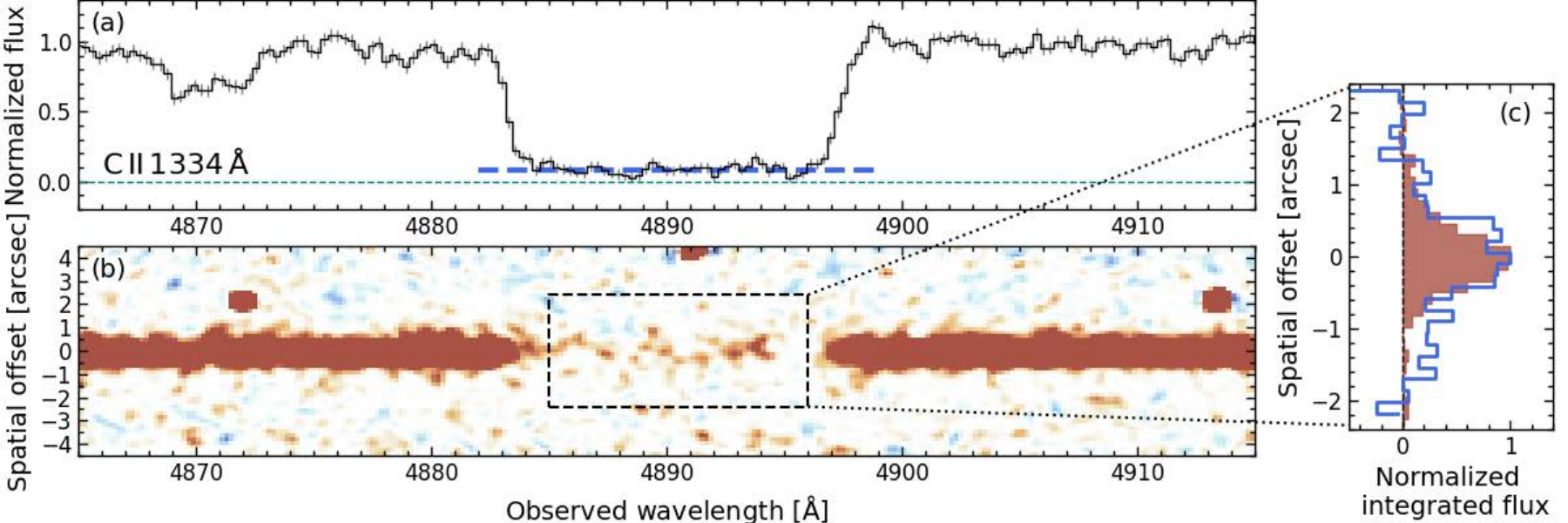} \\
\includegraphics[trim={0.0cm 0.0cm 0.0cm 0.0cm},clip,width=\textwidth]{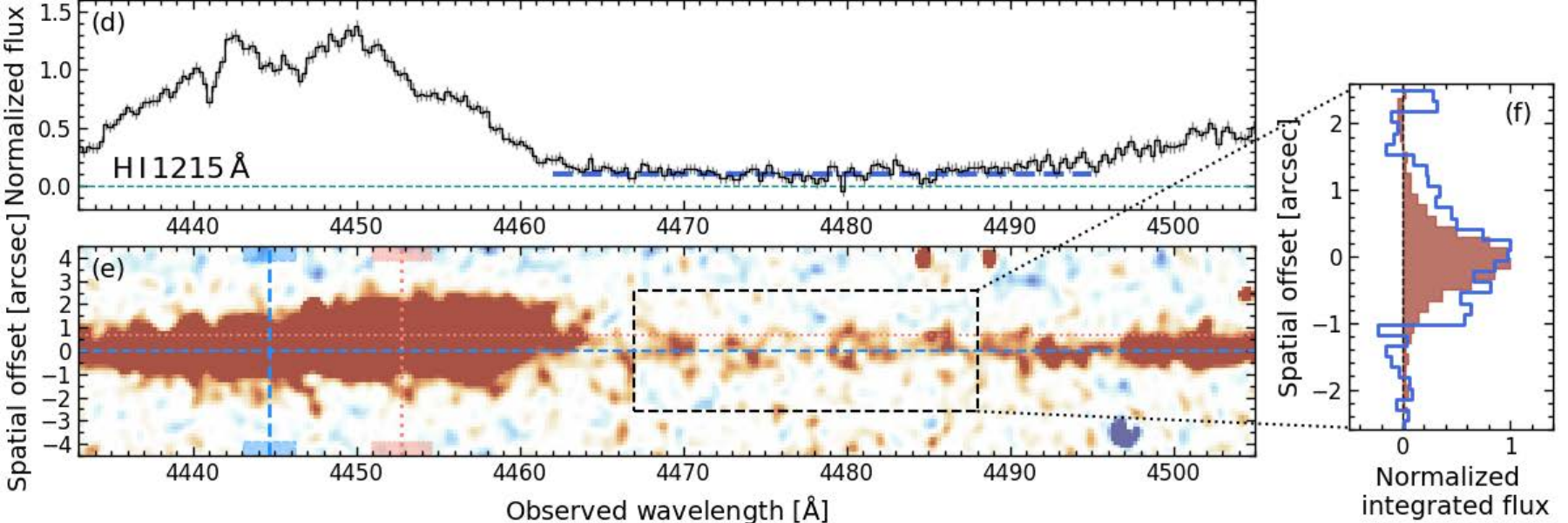}
\caption{
\textbf{Spectral regions around \CII\,1334\,\AA\ and Ly$\alpha$ lines at $z=2.622$.} 
\noindent
\textbf{a}, \textbf{b}, \textbf{c}, \CII\,1334\,\AA\, line. \textbf{d}, \textbf{e}, \textbf{f}, Ly$\alpha$ line. The strong \CII\ and Ly$\alpha$ absorption suppress light from the quasar’s central engine, acting like a coronagraph which reveals the $\sim$10 times fainter emission from the host galaxy as residual emission in the absorption trough. 
\textbf{a}, \textbf{d}, the portions of the 1D X-shooter spectrum extracted along the quasar trace, with cyan and blue dashed lines indicating the zero flux level and the average line flux residual, respectively. The error bars represent the standard deviation of the measured flux. \textbf{b}, \textbf{e}, the 2D X-shooter spectrum (smoothed for illustration purposes) of the same spectral regions. \textbf{c}, \textbf{f}, the spatial profiles integrated along the dispersion axis, with the blue histogram corresponding to regions within the saturated lines (black dashed boxes in \textbf{b}, \textbf{e}), while the red filled histogram correspond to the continuum emission. This faint emission clearly extends beyond the point spread function, in agreement with an origin in an extended object.
In \textbf{e}, the blob at $\lambda\sim4430-4465$\,\AA\ corresponds to the extended Ly$\alpha$ emission seen in the bottom of the extremely strong Ly$\alpha$ absorption line. The red dotted and blue dashed lines denote the centroids of the ALMA CO emission of the companion and quasar host galaxies, respectively in both spatial and velocity space, with the stripe indicating their velocity spread. 
\label{fig:pc_Lya}
}

\end{figure*}

\begin{figure*}
\centering
\includegraphics[trim={0.0cm 0.0cm 0.0cm 0.0cm},clip,width=\textwidth]{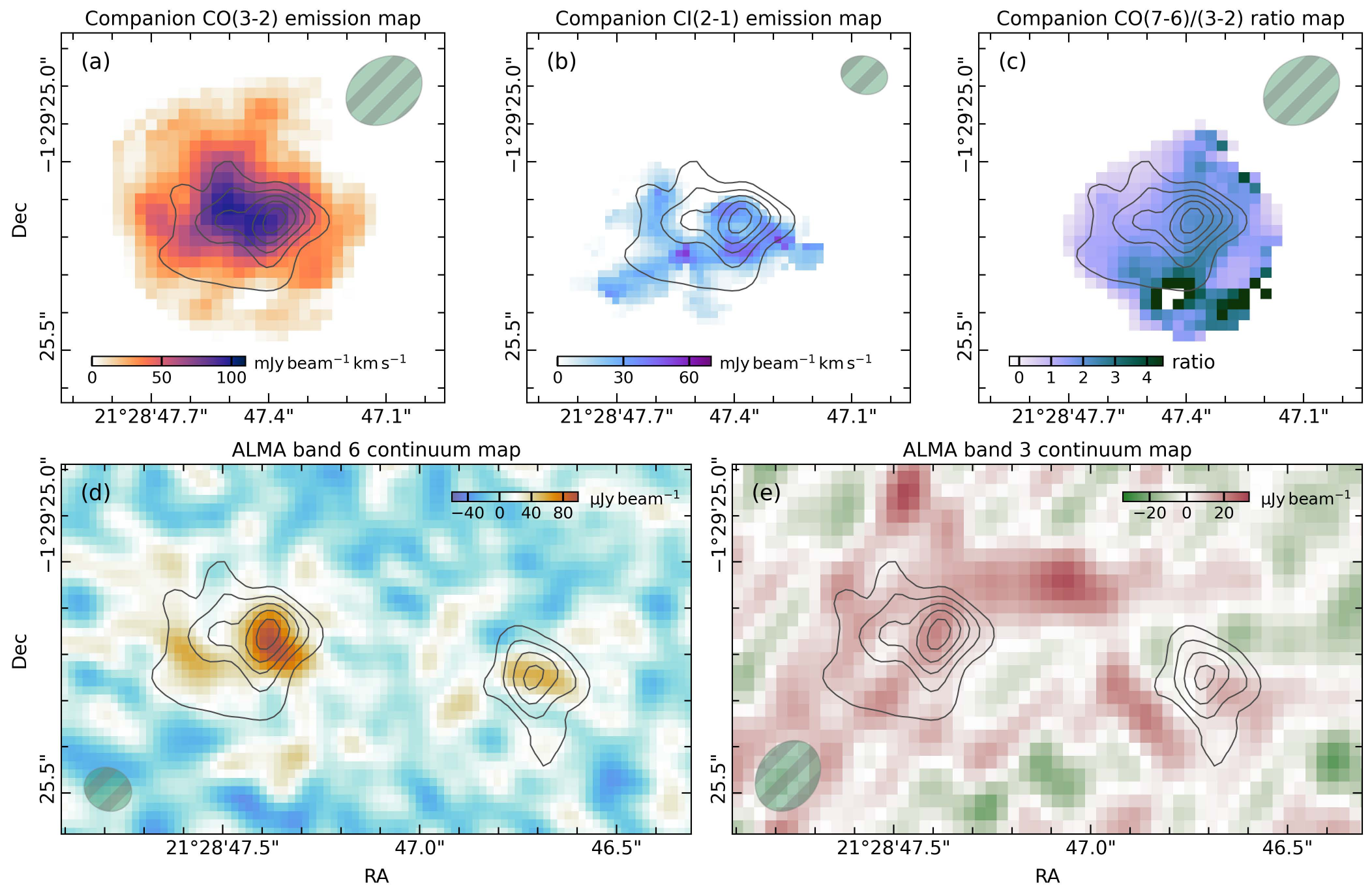}
\caption{
\textbf{Line and continuum ALMA images of the field around \qso.} 
\noindent
\textbf{a}, \textbf{b}, \textbf{c}, The integrated emission of CO(3-2) and \CI(2-1) lines, together with CO(7-6)/(3-2) line ratios, respectively, centered on the companion galaxy. \textbf{d}, \textbf{e}, ALMA band 6 and 3 wideband continuum images, respectively. The gray contours indicate the CO(7-6) emission shown in Fig.~1. The green hatch ellipses in each panel represent the synthesized beam.
\label{fig:alma_add}
}
\end{figure*}

\begin{table}
\caption{
\textbf{Spectral energy distribution fitting of the companion and quasar host galaxy with {\sc bagpipes}.} \label{tab:sedfit}
}
\vspace{-0.4cm}
\begin{tabular}{ccccc}
\hline
\hline
parameter & description & prior & \multicolumn{2}{c}{posterior}  \\
\hline
\multicolumn{3}{c}{} & Companion galaxy & Host galaxy \\
\hline
z & redshift & Gaussian from CO & $2.663\pm0.001$ & $2.656\pm0.001$ \\ 
$\log M_{\rm *}, [M_{\odot}]$ & stellar mass & log10 & $10.84^{+0.12}_{-0.12}$ & $10.7^{+0.3}_{-0.3}$ \\ 
$\alpha_{\rm dpl}$ & falling SFH slope & uniform & $<0.1$ & $-0.8^{+0.8}_{-0.8}$\\
$\beta_{\rm dpl}$ & rising SFH slope & uniform & $>1.5$ & $-2..3$\\
$\tau_{\rm dpl}$ [Gyr] & SFR peak time & log10 & $2.2^{+0.1}_{-0.2}$ & $0.5..2$\\ 
$Z$ & metallicity & Gaussian and log10 & $<0.9$ & $0.8..2.1$\\
A$_{\rm V}$ & extinction & log10 & $1.2^{+0.1}_{-0.1}$ & $0.7^{+0.2}_{-0.3}$ \\
$U_{\rm min}$ & dust parameter & log10 & $>3.2$ & $>1.3$\\
$\gamma$ & dust parameter & log10 & $>0.9$ & $>0.3$\\
$\log U$ & ionization parameter & uniform & $<-3.3$ & $>-2.4$\\ 
\hline
\multicolumn{5}{c}{Derived quantities:}  \\
\hline
SFR $[\rm M_{\odot}\,yr^{-1}]$ & \multicolumn{2}{c}{current star-formation rate} & $250^{+60}_{-50}$ & $24^{+17}_{-11}$\\
\hline
\hline
\end{tabular}
\end{table}

\begin{figure*}
\centering
\includegraphics[trim={0.0cm 0.0cm 0.0cm 0.0cm},clip,width=\textwidth]{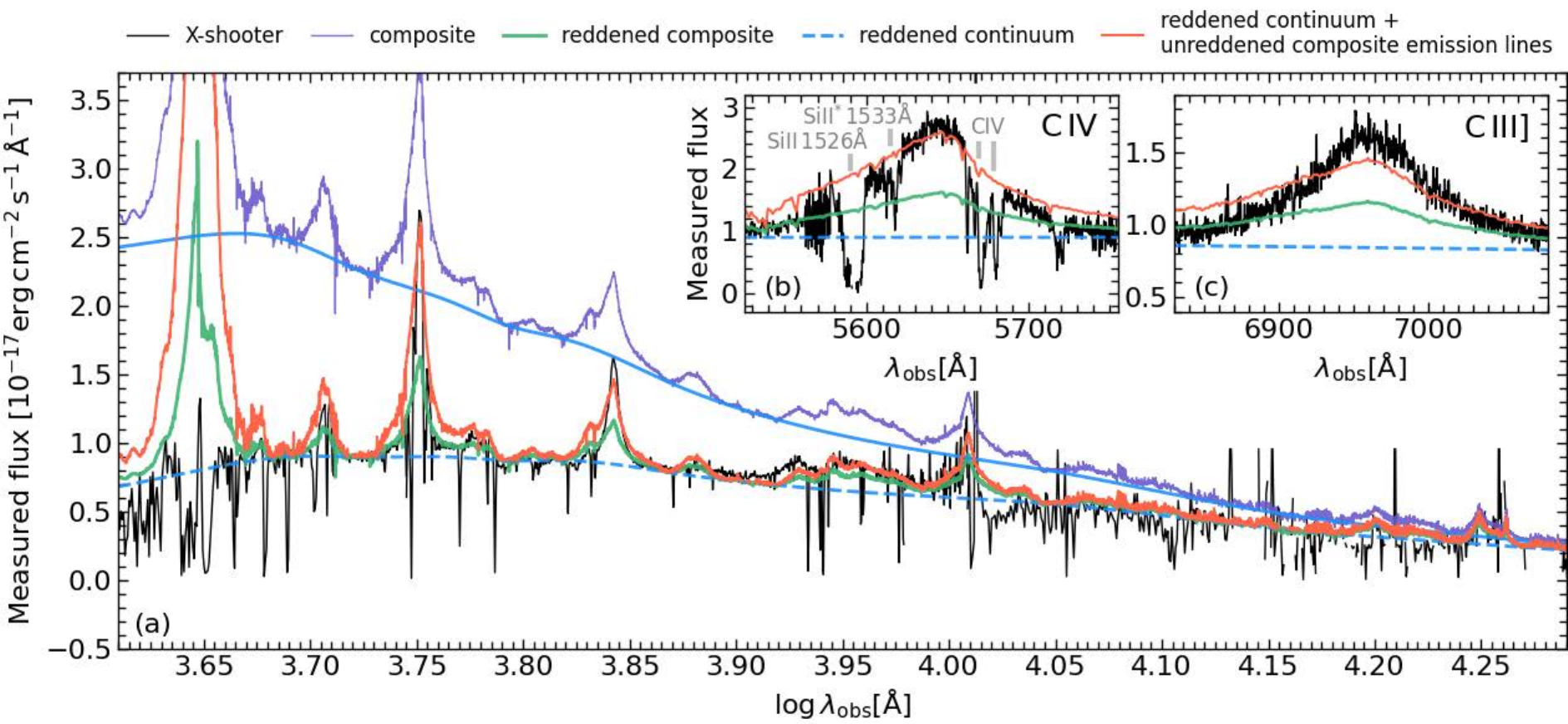}
\caption{
\textbf{A full view of the X-shooter spectrum of \qso.}
\noindent
The black, purple, green lines show respectively the observed spectrum, the quasar composite spectrum from \cite{Selsing2016} and the same composite reddened by an SMC-like extinction law with $A_{\rm V} = 0.2$ at z=2.66. The blue line depicts the continuum component of the quasar spectrum (i.e. removing emission lines) reddened by the same extinction. The red spectrum is the sum of the reddened continuum and the unreddened emission lines and clearly reproduces much better the emission lines redwards of Ly-$\alpha$. This indicates that the continuum light from the quasar accretion disc passes through dusty gas, while light from the emission line regions is almost unaffected, implying that dust is confined in much smaller regions. \textbf{b}, \textbf{c} indicate zoom-in on the spectral region around \CIV\ and \CIII] emission lines. The position of metal absorption lines associated with the proximate DLA are marked by grey ticks and labels.
\label{fig:diff_red}
}
\end{figure*}

\begin{figure*}
\centering
\includegraphics[trim={0.0cm 0.0cm 0.0cm 0.0cm},clip,width=\textwidth]{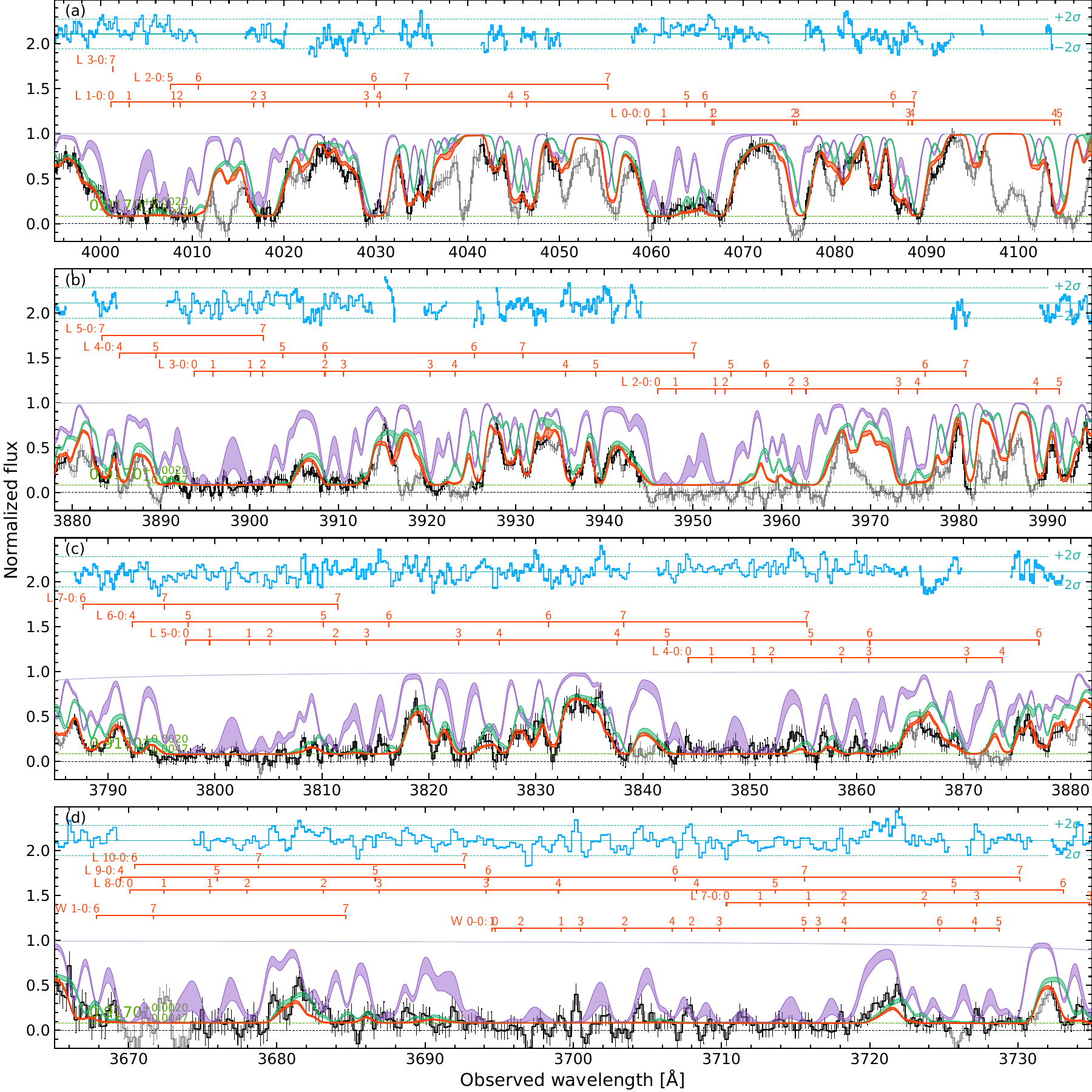}
\caption{
\textbf{Fit to the line profiles of H$_2$ lines associated with the proximate absorber at $z=2.662$ towards \qso.} 
\noindent
The black lines represent the observed normalized spectrum, with the error bars representing the standard deviation. The green, violet, and red stripes indicate the profiles of H$_2$ components $A$, $B$, and the total profile, respectively. Black and green dashed horizontal lines at the bottom of each panel show the zero flux level and partial coverage, respectively. Blue step-like lines at the top of each panel display the residuals, with horizontal lines representing $-2\sigma$, $0$, and $2\sigma$ levels. Red horizontal segments identify the band (Lyman/Werner, vibrational level of upper and lower state) as well as main rotational levels (numbers above ticks, only shown for component $A$ for clarity). Note that lines from low rotational levels in the L2-0 band (around $3950$\,\AA) are blended with the strong Ly$\alpha$ line from intervening \HI\ at $z=2.25$, identified through its associated metal transitions.  
\label{fig:H2_fit}
}
\end{figure*}

\begin{figure*}
\centering
\includegraphics[trim={0.0cm 0.0cm 0.0cm 0.0cm},clip,width=\textwidth]{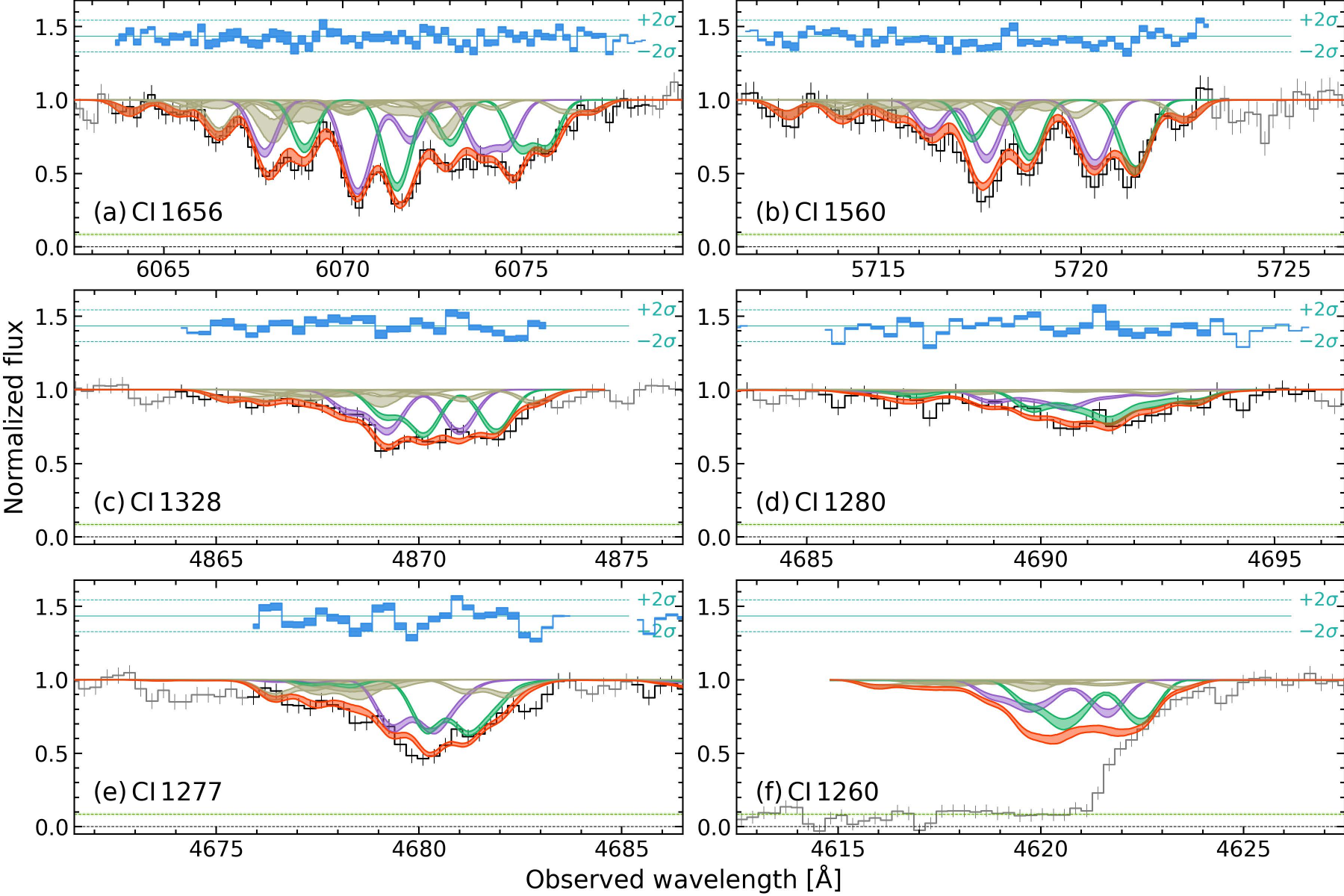}
\caption{
\textbf{Fit to the \CI\ absorption line profiles associated with the proximate absorber at $z\approx2.662$ towards \qso.} 
\noindent
The graphical elements are the same as in Extended Data Fig.~6 with profiles of additional components shown by gray lines. Each profile contains absorption from three fine-structure levels of the ground state.
\label{fig:CI}
}
\end{figure*}

\begin{figure*}
\centering
\includegraphics[trim={0.0cm 0.0cm 0.0cm 0.0cm},clip,width=\textwidth]{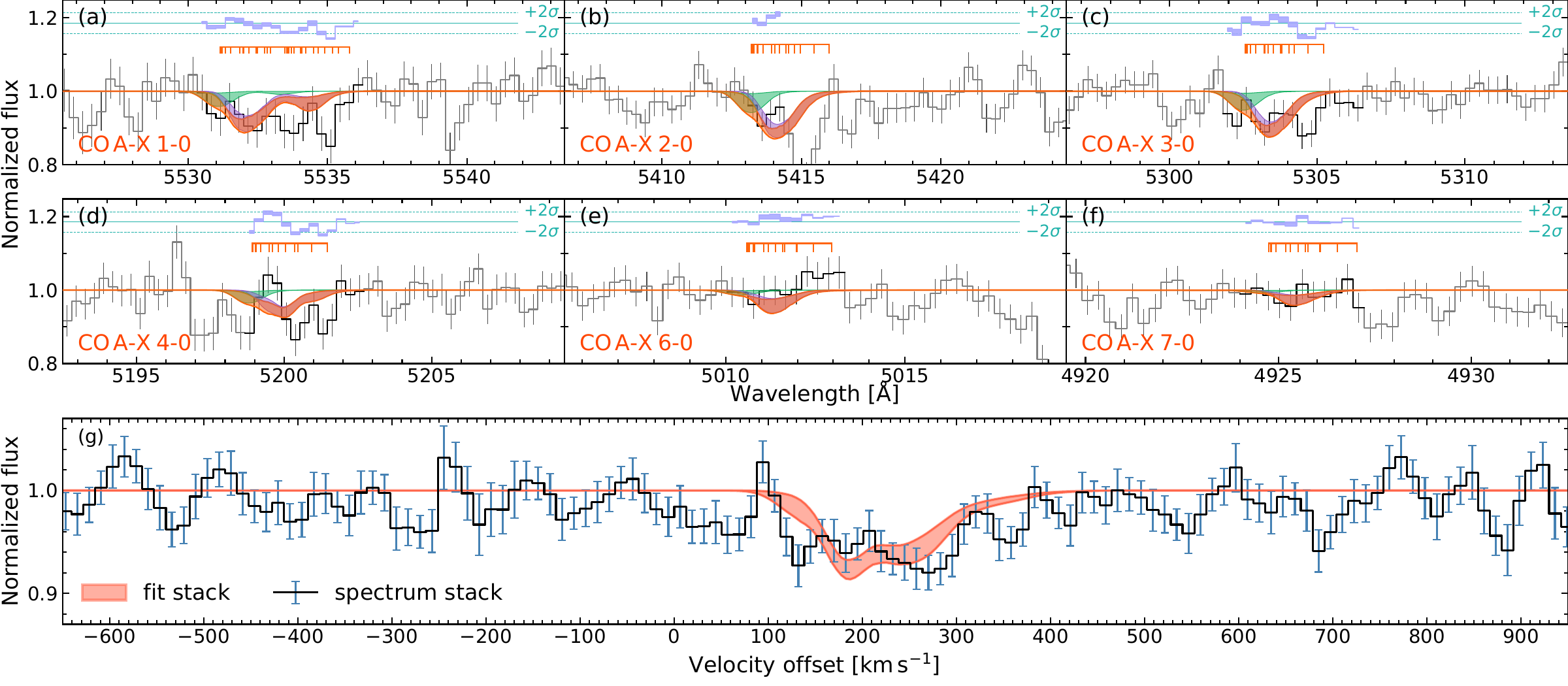}
\caption{
\textbf{Possible detection of CO absorption lines at $z=2.662$ in the proximate absorber towards \qso.}
\noindent
\textbf{a}-\textbf{f}, The fit to individual CO bands. The black line show the observed spectrum, while the green, blue and red lines represent the two individual components and total line profiles. \textbf{g}, The stack of the observed spectrum (black) and total profile model (red).
\label{fig:CO}
}
\end{figure*}

\end{document}